\documentstyle[a4wide,12pt,francais]{article}
\begin{document}
\sloppy
 
\newcommand{\hh}{{\cal H}}
\newcommand{\br}{{\bf R}}
\newcommand{\bc}{{\bf C}}
\newcommand{\bn}{{\bf N}}
\newcommand{\lt}{{L^2 ({\br},dx)}}
\newcommand{\ltp}{{L^2 ({\br},dp)}}
\newcommand{\hs}{{\cal H}}
\newcommand{\st}{{\cal S}(\br)}
\newcommand{\std}{{\cal S}^{\prime} (\br)}
\newcommand{\pr}{\prime}
\newcommand{\pa}{\partial}
\newcommand{\ds}{\displaystyle}
\renewcommand{\dag}{\ds{\dagger}}
 
\newcommand{\ri}{{\rm i}}

\newtheorem{defin}{D\'efinition}
\newtheorem{theo}{Th\'eor\`eme}
\newtheorem{lem}{Lemme}
 
\hfill LYCEN 9960b
\vskip 0.07truecm
\hfill Juillet 1999
%\vskip 0.07truecm 
%\hfill qm ... 

\bigskip

\thispagestyle{empty}
 
\begin{center}
{\bf \Huge{Formalisme de Dirac et surprises math\'ematiques  
en m\'ecanique quantique}\footnote{Ceci est la traduction
fran\c caise du preprint LYCEN 9960a. Ce travail a \'et\'e 
soutenu par la Fondation Alexander von Humboldt lors d'un 
s\'ejour de l'auteur \`a l'Institut f\"ur Theoretische Physik
de l'Universit\'e de G\"ottingen.}}
\end{center}
\bigskip
\bigskip
\bigskip
\centerline{D\'edi\'e \`a la m\'emoire de {\bf Tanguy Altherr}
(1963 - 1994)
\footnote{Je
d\'edie ces notes \`a la m\'emoire de Tanguy Altherr
qui nous a quitt\'e de mani\`ere tr\`es brutale dans
les montagnes qu'il aimait tant (et o\`u j'ai pu faire de belles 
 courses avec lui).
Pour ce qui est du sujet de ces notes (sujet duquel
j'ai eu le plaisir de discuter avec lui),
Tanguy appr\'eciait bien
le formalisme de
Dirac et savait certainement l'appliquer
\`a de vrais probl\`emes physiques (comme ceux sur lesquels
il travaillait avec un enthousiasme,
une \'energie et une productivit\'e
impressionnantes).
M\^eme s'il ne partageait pas
mes pr\'eoccupations \`a ce sujet, il aimait bien discuter
des probl\`emes concernant le formalisme g\'en\'eral
de la physique quantique.}}
\bigskip
\bigskip
\bigskip
\centerline{{\bf Fran\c cois Gieres}}
\bigskip
\centerline{\it Institut de Physique Nucl\'eaire}
\centerline{\it Universit\'e Claude Bernard (Lyon 1)}
\centerline{\it 43, boulevard du 11 novembre 1918}
\centerline{\it F - 69622 - Villeurbanne Cedex}
\bigskip
\bigskip
\bigskip
\bigskip
\begin{center}
{\bf \Large{R\'esum\'e}}
\end{center}
Diff\'erents formalismes
sont utilis\'es en m\'ecanique quantique
pour la description des
\'etats et des observables :
la m\'ecanique ondulatoire, la m\'ecanique matricielle et le formalisme
invariant.
Nous discutons les probl\`emes et 
inconv\'enients du formalisme invariant
ainsi que ceux de la notation des bras et kets
introduite par Dirac dans ce contexte.
Nous indiquons
comment tous les probl\`emes peuvent \^etre r\'esolus ou du moins
\'evit\'es.
Une s\'erie d'exemples 
illustre les points 
soulev\'es et montre comment l'insouciance math\'ematique
peut ais\'ement conduire \`a des contradictions 
math\'ematiques surprenantes. 
 
\newpage
 
\tableofcontents
 
\newpage
 
\setcounter{page}{1}

\section{Introduction}
 
Les diff\'erentes formulations ou `repr\'esentations' utilis\'ees
en m\'ecanique quantique pour
la description des \'etats d'une
particule (ou d'un syst\`eme de particules)
sont essentiellement au nombre de trois :
la m\'ecanique ondulatoire, la m\'ecanique matricielle
et le formalisme invariant.
Les deux premi\`eres utilisent des espaces de Hilbert
concrets et la troisi\`eme
un espace de Hilbert
abstrait. En g\'en\'eral la derni\`ere formulation est
pr\'esent\'ee moyennant la notation
des bras et kets de Dirac \cite{d},
notation qui va d'habitude de pair
avec une certaine interpr\'etation des op\'erations
math\'ematiques donn\'ee par Dirac. Rappelons-en les
principaux ingr\'edients :
\begin{eqnarray*}
& \bullet & \quad
| \Psi \rangle \quad {\rm et} \quad \langle \Psi |
\\
& \bullet & \quad
A | \Psi \rangle \quad {\rm et} \quad \langle \Psi |A^{\dag}
\\
& \bullet & \quad
\langle \Phi |A|\Psi  \rangle
\\
& \bullet & \quad
\left\{ |n \rangle \right\}_{n \in \bn }
\quad {\rm et} \quad
\left\{ |x \rangle \right\}_{x \in \br }
\\
& \bullet & \quad
|n_1, n_2, ... \rangle
\quad \mbox{associ\'e \`a un ECOC}
\quad \left\{ A_1, A_2 , ... \right\}
\ \ .
\end{eqnarray*}
Dans la section prochaine et dans l'annexe A,
nous pr\'eciserons ces notations
tout en rappelant quelques notions math\'ematiques importantes.
Dans les sections 3 et 4,
nous discuterons successivement les questions sui\-vantes :
\begin{enumerate}
\item
Est-ce qu'une repr\'esentation est pr\'ef\'erable \`a une autre
du point de vue math\'ematique ou pratique? En particulier,
nous discuterons le statut du {\em formalisme invariant} auquel
la pr\'ef\'erence est donn\'ee dans la plupart des ouvrages
r\'ecents.
\item
Quels sont les avantages, inconv\'enients et probl\`emes
des {\em notations de Dirac} et de leur interpr\'etation?
(Les r\`egles de calcul d\'eduites de ces notations et de leur 
interpr\'etation sont d'habitude appliqu\'ees dans le cadre
du formalisme invariant et repr\'esentent alors un calcul 
symbolique.) 
\end{enumerate}
 
Pour anticiper nos r\'eponses \`a ces questions,
nous signalons d'ores et d\'ej\`a que nous aboutirons
\`a la conclusion que l'application
{\underline{\em syst\'ematique}} du formalisme invariant
et l'usage {\underline{\em rigide}}
des notations de Dirac - formalisme et usage pr\^on\'es
dans la majorit\'e
des trait\'es modernes de la m\'ecanique quantique -
ne sont pas recommandables ni du point de vue math\'ematique ni du
point de vue pratique. Des compromis qui retiennent les
avantages de ces formalismes tout en \'evitant
leurs inconv\'enients
seront indiqu\'es.  
Les conclusions qu'on peut en tirer pour la pratique 
et pour l'enseignement de
la th\'eorie quantique sont r\'esum\'ees dans la section finale. 
 
Dans ce contexte, il est peut-\^etre utile de
mentionner que la monographie classique de Dirac \cite{d}
(et donc la majorit\'e des ouvrages modernes qu'elle a inspir\'es)
contient un bon nombre d'affirmations qui sont ambigu\"es
ou incorrectes du point de vue math\'ematique : ces points
ont \'et\'e soulev\'es et discut\'es par Jauch \cite{jau}.
L'\'etat des choses peut \^etre d\'ecrit de la mani\`ere 
suivante \cite{grau} : ``L'\'el\'egance, la clart\'e apparente 
et la force percutante du formalisme de Dirac sont 
malheureusement acquises au d\'etriment de l'introduction de 
fictions math\'ematiques. [...] On a une `machinerie' formelle
dont la signification est imp\'en\'etrable, surtout pour le d\'ebutant, 
et dont celui-ci ne peut pas reconna\^{i}tre la 
probl\'ematique."
Aussi le verdict de math\'ematiciens majeurs comme J.Dieudonn\'e
est foudroyant \cite{dieu} : 
``Mais c'est lorsqu'on aborde les th\'eories math\'ematiques 
qui sont \`a la base de la m\'ecanique quantique que l'attitude 
de certains physiciens dans le maniement de ces th\'eories 
confine v\'eritablement au d\'elire. [...] On se demande ce qui peut
rester dans l'esprit d'un \'etudiant 
lorsqu'il a absorb\'e cette invraisemblable accumulation de non-sens,
une v\'eritable `bouillie pour les chats'! Ce serait \`a croire que les 
physiciens d'aujourd'hui ne sont \`a l'aise que dans le flou, 
l'obscur et le contradictoire.''
Certes nous pouvons reprocher \`a beaucoup de math\'ematiciens 
leur intransigeance et leur refus de faire le moindre effort pour 
comprendre des formulations manquant de rigueur,  
mais leur jugement devrait quand m\^eme nous donner \`a r\'efl\'echir,
une r\'eflexion \`a laquelle 
nous esp\'erons contribuer de mani\`ere constructive 
par le pr\'esent travail.

En fait, on ne peut pas nier que 
l'insouciance math\'ematique (qui est pratiquement inh\'erente 
au calcul symbolique de Dirac) 
conduit souvent et vite \`a des contradictions 
apparentes qui sont parfois tr\`es \'etonnantes : 
 nous allons illustrer ceci par une s\'erie d'exemples simples 
 qui sont pr\'esent\'es
dans l'annexe B. 
Dans la litt\'erature, de telles contradictions se sont manifest\'ees
dans l'\'etude de ph\'enom\`enes physiques plus
compliqu\'es et elles ont m\^eme conduit \`a la mise en question
d'effets physiques comme par exemple
l'effet d'Aharonov et Bohm \cite{ab}.
Les contradictions
peuvent seulement \^etre \'ecart\'ees en faisant appel \`a
une formulation math\'ematique
plus pr\'ecise des probl\`emes, formulation qui va souvent de pair
avec une compr\'ehension physique plus profonde des ph\'enom\`enes
\'etudi\'es.
Dans les annexes A et C, nous introduisons les outils math\'ematiques 
appropri\'es (qui sont bien connus en physique 
math\'ematique, mais ne pas \'evoqu\'es dans la plupart des ouvrages  
de m\'ecanique quantique) et nous montrons
comment ils permettent de r\'esoudre de mani\`ere efficace tous les
probl\`emes pos\'es.

\section{M\'ecanique quantique et espaces de Hilbert}
 
Les diff\'erentes repr\'esentations
utilis\'ees en m\'ecanique quantique
sont discut\'ees dans de nombreux ouvrages
\cite{ct}
et nous les r\'esumerons
dans la suite pour fixer les notations. La th\'eorie
math\'ematique sous-jacente est
expos\'ee dans les livres d'analyse
fonctionnelle \cite{af}.
Parmi ceux-ci il existe d'excellentes
monographies qui pr\'esentent la th\'eorie g\'en\'erale
avec
ses applications \`a la m\'ecanique quantique
\cite{rs,sg,bgc} (voir aussi \cite{ri,lio,tj,krey}).
 
\subsection{Les diff\'erents espaces de Hilbert}
 
Nous consid\'erons le mouvement
d'une particule sur une droite
param\'etr\'ee par $x \in {\bf R}$. (La g\'en\'eralisation
\`a un intervalle born\'e, \`a $3$ dimensions,
au spin ou \`a un syst\`eme de particules ne pose pas de
probl\`emes notables.)
Les espaces de Hilbert utilis\'es pour
la description des \'etats de la
particule sont essentiellement au nombre de trois.
 
%\bigskip
 
 \newpage 
 
\noindent
{\bf (1) ``M\'ecanique ondulatoire" :} [de Broglie, Schr\"odinger,
1923 - 1926]
 
On consid\`ere
l'espace des fonctions de carr\'e sommable (fonctions d'onde),
$$
\lt = \{ f : {\bf R} \to {\bf C} \ \mid \ \int_{{\bf R}} dx \,
| f(x) |^2 < \infty \}
\ \ ,
$$
avec le produit scalaire\footnote{Le complexe conjugu\'e
de $z \in {\bf C}$ est not\'e par $\bar z$ ou $z^{\ast}$.}
$$
\langle
f, g \rangle _{L^2} = \int_{{\bf R}} dx \; \overline{f(x)} \, g(x)
\qquad \mbox{pour} \ \; f,g \in \lt
\ \ .
$$
Cet espace est reli\'e
par la transformation de Fourier
\`a l'espace de Hilbert
$L^2 ({\bf R} , dp )$ des fonctions d'onde d\'ependant
de l'impulsion $p$ :
\begin{eqnarray}
\label{f}
{\cal F} \ : & \lt & \longrightarrow \ \ L^2 ( {\br}, dp)
\\
 & f & \longmapsto \ \  {\cal F} f
\quad {\rm avec} \ \;
\left( {\cal F} f \right) (p)
= {1 \over \sqrt{2\pi \hbar}} \int_{\br} dx \, f(x) \,
{\rm exp} (-{\ri \over \hbar}px )
\ \ .
\nonumber
\end{eqnarray}

\noindent {\bf (2)
``M\'ecanique matricielle" :} [Heisenberg, Born, Jordan, Dirac,
1925 - 1926]
 
On travaille avec
l'espace des suites infinies de carr\'e sommable,
$$
l_2 = \{ \vec{x} = (x_1, x_2, ...) \mid x_k \in {\bf C}
\ \ {\rm et} \ \; \sum_{k=1}^{\infty} \ | x_k | ^2 < \infty \}
\ \ ,
$$
avec le produit scalaire
$$
\langle \vec x , \vec y \, \rangle _{l_2} = \sum_{k=1} ^{\infty}
\overline{x_k} \, y_k
\ \ .
$$
 
L'\'equivalence entre la m\'ecanique ondulatoire et la m\'ecanique
matricielle a \'et\'e d\'emontr\'ee en 1926 par Schr\"odinger.
Elle repr\'esentait le point de d\'epart pour la recherche
d'une version ``invariante" de la m\'ecanique quantique,
recherche qui conduisit,
par l'interm\'ediaire
des travaux de Dirac et Jordan, \`a l'\'etude des op\'erateurs
lin\'eaires agissant sur un espace de Hilbert abstrait \cite{fh,jvn}.

 Au fait, 
l'espace $l_2$ a \'et\'e introduit en 1912 par D.Hilbert 
dans ses travaux sur les \'equations int\'egrales,
mais une d\'efinition axiomatique de l'espace de Hilbert 
a seulement \'et\'e donn\'ee en 1927 par J.von Neumann 
dans un travail sur le fondement math\'ematique de la m\'ecanique
quantique \cite{krey}. 
C'est une co\"\i ncidence remarquable que la monographie de Courant et
Hilbert \cite{ch} d\'eveloppant les math\'ematiques  de l'espace de Hilbert
ait paru en 1924 et qu'elle semble avoir \'et\'e \'ecrite express\'ement
pour les
physiciens de l'\'epoque\footnote{D.Hilbert :
``I developed my theory of
infinitely many variables from purely mathematical interests
and even called it `spectral analysis' without any
pressentiment that it would later find an application
to the actual spectrum of physics." \cite{cr}}.
Dans la suite, cette th\'eorie a trouv\'e des raffinements
qui sont principalement
d\^us \`a von Neumann \cite{jvn}, \`a Schwartz \cite{ls} et
\`a Gelfand \cite{gv} et qui
permettent de d\'ecrire d'une mani\`ere pr\'ecise toutes les observables
et \'etats en m\'ecanique quantique.
 
\bigskip

%\newpage 

\noindent {\bf (3)
``Formalisme invariant" :} [Dirac, Jordan, von Neumann,
1926 - 1931]
 
On utilise
un espace de Hilbert complexe abstrait $\hs$
qui est {\em s\'eparable}
(ce qui veut dire qu'il admet une base orthonorm\'ee
constitu\'ee d'une famille d\'enombrable de vecteurs) et
{\em de dimension infinie}. 

Dans l'annexe A, nous avons
r\'esum\'e les
notions fondamentales de la th\'eorie des espaces de Hilbert
ainsi que la notation des bras et kets
d\'evelopp\'ee par Dirac dans ce cadre 
\`a partir de 1939 \cite{d}.
Dirac a invent\'e ces notations astucieuses
suite \`a une interpr\'etation
parti\-culi\`ere des expressions faisant intervenir les vecteurs et
op\'erateurs.
L'interpr\'etation qu'il en a donn\'e
et les avantages et inconv\'enients qui en r\'esultent
seront discut\'es dans la section 4.

\subsection{Liens entre les espaces de Hilbert}
 
Pour d\'ecrire les liens entre les espaces
de Hilbert introduits dans la section pr\'ec\'edente,
nous avons besoin des concepts d'op\'erateur
unitaire et d'isomorphisme \cite{rs}.
 
\begin{defin}
Pour $i=1,2$, soit $\hs_i$ un espace de Hilbert complexe s\'eparable 
avec produit
scalaire $\langle \ , \ \rangle _{\hs_i}$. Un op\'erateur lin\'eaire
$U  : \hs_1 \to \hs_2$ est dit {\em unitaire} si
 
\noindent {\rm (i)} $U$ est partout d\'efini sur $\hs_1$.
 
\noindent {\rm (ii)} L'image de $\hs_1$ par $U$ est $\hs_2$ tout entier.
 
\noindent {\rm (iii)} $U$ pr\'eserve le produit scalaire :
\begin{equation}
\label{un}
\langle U f , U g \rangle _{\hs_2} = \langle f , g \rangle _{\hs_1}
\qquad \mbox{pour tout} \quad f,g \in \hs_1
\ \ .
\end{equation}
 
Deux espaces
de Hilbert $\hs_1$ et $\hs_2$ qui sont reli\'es par un op\'erateur
unitaire sont dits {\em isomorphes} et on \'ecrit
$\hs_1 \simeq \hs_2$.
\end{defin}
 
Concernant les espaces de Hilbert intervenant en m\'ecanique quantique,
nous disposons
d'un r\'esultat classique de l'analyse fonctionnelle :
\begin{theo}
{\rm (i)} Les espaces de Hilbert complexes $l_2, \,  \lt$ et $\ltp$
sont s\'epa\-rables et de dimension infinie.
 
\noindent
{\rm (ii)} Tout espace de Hilbert complexe qui est s\'eparable et de
dimension infinie est iso\-morphe \`a $l_2$.
\end{theo}
 
Il suit de ce
r\'esultat que tous les espaces de Hilbert introduits
ci-dessus sont isomorphes :
\begin{equation}
\fbox{\mbox{$
\hs \simeq l_2 \simeq \lt \simeq \ltp$}}
\ \ .
\end{equation}
En particulier,
le th\'eor\`eme de Parseval et Plancherel dit que
la transformation de Fourier (\ref{f}) r\'ealise
l'isomorphisme entre $\lt$ et $\ltp$:
\[
\langle {\cal F} f, {\cal F} g \rangle = 
\langle  f,  g \rangle
\qquad \mbox{pour tout} \ \; f,g \in \lt
\ \ .
\]

D'une mani\`ere g\'en\'erale,
le passage entre $\hs$ et les autres espaces se fait
en choisissant une base orthonorm\'ee 
$\{ | n \rangle \}_{n \in \bn}$ 
(ou
une base orthonorm\'ee g\'en\'eralis\'ee
$\{ | x \rangle \}_{x \in \br}$) de $\hs$ et en associant \`a tout
vecteur  $| \Psi \rangle \in \hs$ l'ensemble de ses composantes par rapport
\`a cette base:
\begin{eqnarray*}
\psi_n & \! \! :=  \! \! & \langle n | \Psi \rangle 
\quad {\rm pour} \ \; n \in \bn 
\qquad , \quad \vec{\psi} \equiv  
(\psi_0 , \psi_1, ...)  \in l_2 
\\
\psi (x) & \! \! := \! \! & \langle x | \Psi \rangle 
\quad {\rm pour} \ \; x \in \br 
\qquad \, , \quad \psi \in \lt 
\ \ .
\end{eqnarray*}
Dans la deuxi\`eme expression, l'action de $\langle x |$ sur  
$| \Psi \rangle$ est \`a comprendre au sens d'une action 
de distribution - voir annexe A.3. 

  Le passage entre $\lt$ et $l_2$ se fait de mani\`ere analogue: 
   \`a la fonction $\psi \in \lt$ on associe la suite 
   $(\psi_0 , \psi_1, ...)  \in l_2 $ constitu\'ee des composantes 
$\psi_n :=  \langle \varphi_n , \psi \rangle_{L^2}$ 
de $\psi$ 
par rapport \`a une base orthonorm\'ee $\{ \varphi_n \}_{n\in \bn}$
de $\lt$.

\section{Discussion du formalisme invariant}

Comme les diff\'erents espaces de Hilbert utilis\'es
en m\'ecanique quantique sont tous isomorphes, ils sont
compl\`etement \'equivalents du point de vue math\'ematique.
(Ils repr\'esentent des r\'ealisations diff\'erentes
d'une m\^eme structure abstraite.)
Cependant, du point de vue pratique, certains
espaces sont plus appropri\'es que d'autres\footnote{G.Orwell :
``All animals are equal, but some animals are more equal
than others."}.
\begin{enumerate}
\item
Le calcul matriciel bas\'e
sur l'espace $l_2$ n'est pas facilement
maniable et ce formalisme n'a gu\`ere \'et\'e utilis\'e
apr\`es les d\'ebuts de la m\'ecanique quantique
(1926) o\`u il a jou\'e
un r\^ole important \cite{vdw}.
\item
L'ar\`ene des ph\'enom\`enes
physiques est l'espace dit de confi\-guration param\'etr\'e
par $x$ et les conditions aux limites ou de r\'egularit\'e
concernent directement les fonctions d'onde d\'efinies sur cet espace,
ce qui favorise l'utilisation de l'espace
de Hilbert $\lt$.
\item
Le choix d'un espace de Hilbert abstrait $\hs$
est d'habitude motiv\'e
par l'analogie avec la g\'eom\'etrie dans l'espace euclidien
$\br^n$ (ou ${\bf C}^n$) : l'utilisation de ``vecteurs abstraits" est
plus g\'eom\'etrique que celle de leurs composantes. Ainsi il est
tentant de travailler avec des vecteurs $ | \Psi \rangle \in \hs$
tout en interpr\'etant les suites de $l_2$ ou les
fonctions de $\lt$ comme les composantes des vecteurs $|\Psi \rangle $
par rapport \`a diff\'erentes bases de $\hs$. Dans cet esprit,
l'utilisation d'un espace de
Hilbert abstrait en m\'ecanique quantique est souvent pr\'esent\'ee
comme \'etant plus g\'en\'erale que la m\'ecanique ondulatoire
ou matricielle \cite{ct}.
Cependant il y a des diff\'erences importantes entre
les espaces vectoriels de dimension finie et ceux de dimension
infinie qui font en sorte que
l'analogie avec la g\'eom\'etrie ordinaire est tr\`es
d\'elicate et douteuse.
Nous allons maintenant
discuter les probl\`emes r\'esultants qui montrent que le choix
d'un espace de Hilbert abstrait en m\'ecanique quantique obscurcit
et complique
des points importants de la th\'eorie.
\end{enumerate}
 
\subsection{Probl\`emes}
 
\begin{itemize}
\item Pour l'\'etude de probl\`emes simples comme
par exemple la d\'etermination du spectre d'\'energie de
l'oscillateur harmonique (pour lequel les fonctions
propres de l'hamiltonien sont des \'el\'ements bien d\'efinis de $\lt$),
il faut commencer par introduire les distributions propres
$|x \rangle$ de l'op\'erateur de position qui n'appartiennent pas \`a
l'espace de Hilbert $\hs$ (annexe A.3).
\item
Comme nous l'avons soulign\'e dans l'annexe A et illustr\'e
dans l'annexe C,
la d\'efinition d'un op\'erateur
lin\'eaire sur un espace de Hilbert de dimension infinie
n\'ecessite la donn\'ee d'une prescription d'op\'eration
\underline{et} d'un domaine de d\'efinition de cette
op\'eration.
Cet aspect ne constitue pas simplement un point de d\'etail
math\'ematique, puisque le spectre de l'op\'erateur
est tr\`es sensible au domaine de d\'efinition
(conditions aux limites, ...).
Par exemple, suivant le choix du domaine de d\'efinition,
le spectre de l'op\'erateur d'impulsion
$\ds{\hbar \over \ri} \ds{d \over dx}$ sur 
un intervalle compact $[a,b] \subset \br$ peut \^etre vide,
tout ${\bf C}$ ou un sous-ensemble de $\br$
(voir annexe C et r\'ef\'erence \cite{sg}).
Alors que ce probl\`eme
est bien pos\'e d\`es le d\'epart pour les fonctions d'onde
d\'efinies sur l'espace de configuration, il ne l'est
pas de la m\^eme fa\c con pour un espace de Hilbert
abstrait.
 
Cette probl\'ematique se retrouve en m\'ecanique statistique
quantique : par exemple, la d\'efinition de la pression
associ\'ee \`a un ensemble de particules confin\'ees dans
un r\'eservoir fait intervenir les conditions aux limites
\cite{rob}.
\item
Dans le formalisme invariant de la m\'ecanique quantique,
une {\em observable} est d\'efinie comme \'etant un
``op\'erateur hermitien dont les vecteurs propres
orthonorm\'es d\'efinissent une base de l'espace de
Hilbert" \cite{ct}. Partant de cette d\'efinition,
on peut alors montrer d'une mani\`ere formelle que les op\'erateurs
de position et d'impulsion sur $\br$ 
sont des observables. 
Des complications notables apparaissent d\'ej\`a dans 
$\br^2$ ou $\br^3$, si l'on consid\`ere des coordonn\'ees 
non-Cart\'esiennes ; par exemple, la composante radiale $P_r$ de 
l'op\'erateur d'impulsion dans $\br^3$
est hermitien, mais elle ne repr\'esente pas une observable 
- voir Messiah \cite{ct} chap.9. 
Et il existe des op\'erateurs nettement plus compliqu\'es comme
par exemple des hamiltoniens avec des
potentiels al\'eatoires, des potentiels du type $1/  x^n$ ou
$\delta_{x_0} (x)$ ou encore des hamiltoniens sur des espaces
de configuration qui ne sont pas topologiquement triviaux
comme pour l'effet de Aharonov et Bohm ou pour les anyons
\cite{rs, bs, ber}.
La d\'efinition d'une observable rappel\'ee plus haut n\'ecessite alors
d'imposer
des conditions {\em ad hoc} sur les fonctions d'onde associ\'ees
aux \'etats propres
(conditions de r\'egularit\'e, de finitude, d'univaluation,
...); par ailleurs, elle n\'ecessite la d\'etermination explicite d'un
syst\`eme orthonorm\'e de vecteurs propres et la v\'erification
de la relation de fermeture pour ce syst\`eme.
 
Dans une approche qui tient compte des domaines de d\'efinition,
une obser\-vable est simplement donn\'ee par un op\'erateur qui est
auto-adjoint (annexe A.2). Cette condition assure que
le spectre de l'op\'erateur est r\'eel et que ses vecteurs propres
(g\'en\'eralis\'es)
engendrent une base (g\'en\'eralis\'ee)
de l'espace de Hilbert (``th\'eor\`eme
spectral de Hilbert").
Par ailleurs, il existe des crit\`eres simples pour d\'ecider
quand un op\'erateur donn\'e est auto-adjoint ou pour classifier
les diff\'erentes mani\`eres suivant lesquelles on peut le
rendre auto-adjoint - voir \cite{rs,sg,th} et annexe C. 
(En g\'en\'eral,
si un op\'erateur admet plusieurs extensions auto-adjointes,
alors ces derni\`eres d\'ecrivent diff\'erentes situations
physiques \cite{aw,rs}.) En particulier, il n'est pas
n\'ecessaire de faire appel \`a des propri\'et\'es {\em ad hoc} de la
fonction d'onde comme celles mentionn\'ees plus haut ou d'essayer
de d\'eterminer un syst\`eme complet de vecteurs propres
orthonorm\'es.
L'int\'er\^et d'une approche simple et pr\'ecise
appara\^it aussi dans la th\'eorie des perturbations \cite{th}
ou la th\'eorie de la diffusion \cite{amj}.
\item
Une notion importante de la m\'ecanique quantique est celle
d'un ECOC (ensemble complet
d'observables qui commutent).
Elle fait intervenir la commutativit\'e d'op\'erateurs
auto-adjoints qui repr\'esente un point d\'elicat pour les
op\'erateurs non born\'es. En effet, deux op\'erateurs
auto-adjoints $A$ et $B$ commutent si et seulement si
tous les projecteurs intervenant dans leurs d\'ecompositions
spectrales respectives commutent \cite{rs}. Malheureusement des
contre-exemples
montrent que pour la commutativit\'e de $A$ et $B$ il n'est
pas suffisant que $[A,B] =0$ sur un sous-espace dense de $\hs$
sur lequel cette relation est bien d\'efinie \cite{rs}.
Certes ces exemples n'apparaissent gu\`ere en pratique,
mais dans une approche qui tient compte des
domaines de d\'efinition,
on dispose de tous les \'el\'ements auxquels il faut
faire appel si une complication math\'ematique se pr\'esente.
\end{itemize}
Concernant les points math\'ematiques soulev\'es, nous
soulignons qu'en m\'ecanique quantique une formulation pr\'ecise
n'est pas seulement n\'ecessaire pour d\'ecider de l'existence
ou de la non-existence d'effets physiques (comme par exemple
l'effet de Aharonov et Bohm \cite{ab}), mais aussi pour
discuter les probl\`emes difficiles de l'interpr\'etation
(th\'eorie de la mesure, objectivit\'e et r\'ealit\'e, ...)
\cite{au}. Par ailleurs,
une telle formulation 
s'applique directement \`a d'autres domaines de la physique,
un exemple \'etant
le chaos dans les syst\`emes dynamiques classiques \cite{pri}.
 
\subsection{``Solution" des probl\`emes}
 
Certains des probl\`emes soulev\'es dans la section pr\'ec\'edente
sont tellement profonds qu'il semble plus opportun de les
\'eviter que d'y rem\'edier. Les complications
pro\-viennent surtout
du fait que l'on veut -
pour des raisons conceptuelles -
mettre en avant la structure g\'eom\'etrique
d'espace de Hilbert qui est sous-jacente \`a la th\'eorie.
Or celle-ci est \'egalement implicite en
m\'ecanique ondulatoire o\`u les probl\`emes soulev\'es
sont {\em absents} ou du moins {\em bien pos\'es d\`es le d\'epart}.
Il est donc facile d'\'eviter les ennuis math\'ematiques
ou du moins de les rendre plus transparents. 

En particulier,
pour l'enseignement  de la m\'ecanique quantique,
une ``solution" \'evidente des probl\`emes
est de donner une introduction \`a la m\'ecanique ondulatoire qui
souligne
les structures g\'eom\'etriques sous-jacentes et d'indiquer
l'arbitraire de cette formulation en passant \`a d'autres
repr\'esentations comme la m\'ecanique matricielle.
En modifiant la d\'efinition explicite
de l'espace de Hilbert et de son produit scalaire,
le formalisme de la m\'ecanique ondulatoire sur $\br$
se g\'en\'eralise ais\'ement
\`a plusieurs dimensions spatiales, au spin ou \`a des
syst\`emes de particules.
(L'arbitraire de la repr\'esentation
peut aussi \^etre mis en \'evidence en discutant
le passage entre $\hs$ et $\lt$
tout en travaillant avec les fonctions
d'onde pour le reste.)

\section{Discussion des notations de Dirac}
 
Comme il a \'et\'e mentionn\'e dans l'introduction, le formalisme
des bras et kets de Dirac se r\'esume d'une part dans une
certaine \'ecriture des vecteurs, formes lin\'eaires, ...
et d'autre part dans une interpr\'etation particuli\`ere
des op\'erations math\'ematiques qui font intervenir ces quantit\'es.

\subsection{Inconv\'enients}
 
Cette  \'ecriture, ou plut\^ot son interpr\'etation,
pr\'esente un certain nombre
d'inconv\'enients plus ou moins g\^enants.
Parmi ceux-ci le plus grave est
le fait qu'il est {\em impossible} de donner un sens pr\'ecis
\`a l'adjoint $A^{\dag}$ (d'un
op\'erateur non born\'e $A$), si l'on adh\`ere strictement \`a
l'interpr\'etation de Dirac (voir \cite{grau} et aussi \cite{fano}).
Rappelons
\`a ce sujet la d\'efinition fondamentale de Dirac (voir par
exemple les \'equations (B.45)
et (B.51) du chap.II de Cohen-Tannoudji et al. \cite{ct}) :
\begin{equation}
\label{cohe}
\left( \langle \Phi | A \right) | \Psi \rangle =
\langle \Phi | \left( A  | \Psi \rangle \right) \equiv
\langle \Phi |  A  | \Psi \rangle
\qquad {\rm avec} \quad
\langle \Phi | A  = \langle A^{\dag} \Phi |
\ \ .
\end{equation}
D'apr\`es ces relations,
on ne sait pas s'il faut
interpr\'eter l'expression
$\langle \Phi |  A  | \Psi \rangle$ comme\footnote{${\cal D}(A)$
 d\'enote le domaine de d\'efinition de l'op\'erateur $A$ (annexe A.1).}
\[
\langle  A^{\dag} \Phi | \Psi \rangle
\qquad
(\, \mbox{dans quel cas} \ \; |\Phi \rangle \in {\cal D} (A^{\dag})
\ \; {\rm et} \ \; | \Psi \rangle \in \hs \, )
\]
ou comme
\[
\langle \Phi | A   \Psi \rangle
\qquad
(\, \mbox{dans quel cas} \ \; |\Psi \rangle \in {\cal D} (A) \
\; {\rm et} \ \; | \Phi \rangle \in \hs \, ),
\]
sauf si l'on r\'eintroduit les parenth\`eses (ce qui enl\`eve
\'evidemment la simplicit\'e et \'el\'egance du calcul).
H\'elas \cite{rs}, il se peut que ${\cal D}(A^{\dag}) = \{ 0 \}$
pour un op\'erateur $A$ d\'efini sur un 
sous-espace dense de $\hs$. M\^eme si ce
cas de figure ne se pr\'esente gu\`ere en pratique,
les exemples 3 et 7 pr\'esent\'es dans l'annexe B 
montrent que l'ignorance des domaines de
d\'efinition peut ais\'ement conduire \`a des contradictions et
r\'esultats incorrects : le traitement correct d'un
probl\`eme
faisant intervenir des op\'erateurs qui ne peuvent pas \^etre partout
d\'efinis (op\'erateurs non born\'es) est donc d\'elicat.
 
Si l'on convient que
$\langle \Phi |  A  | \Psi \rangle$ est \`a interpr\'eter comme
$\langle \Phi | \left( A  | \Psi \rangle \right)$ 
de sorte que 
$\langle \Phi |A$ ne repr\'esente pas 
$\langle A^{\dag} \Phi |$, mais 
simplement une composition d'applications
(selon l'\'equation (\ref{copo})), alors les ambigu\"{i}t\'es
math\'ematiques concernant les \'el\'ements matriciels sont 
\'ecart\'ees. Cependant il reste encore des inconv\'enients : nous allons
discuter ceux-ci dans le cas habituel o\`u l'on applique 
strictement la notation des bras et kets dans
un espace de Hilbert abstrait $\hs$ qui est de dimension infinie.
 
\begin{itemize}
\item {\em Notation rigide :}
Rappelons d'abord
la d\'efinition standard de l'adjoint $A^{\dag}$
d'un op\'erateur lin\'eaire\footnote{Pour \'eviter la discussion des
domaines de d\'efinition, nous supposons que $A$
est un op\'erateur born\'e : $A$ et $A^{\dag}$
peuvent alors \^etre d\'efinis sur tout l'espace $\hs$.}
$A : \hs \to \hs$ :
\begin{equation}
\label{ad}
\langle \, | \Phi \rangle \, , \, A | \Psi \rangle \, \rangle _{\hs}
\, = \,
\langle \, A^{\dag} | \Phi \rangle \, ,\, |\Psi \rangle \, \rangle _{\hs}
\qquad 
\mbox{pour tout} \ \, |\Phi \rangle , |\Psi \rangle \in \hs 
\ \ .
\end{equation}
Si l'on adh\`ere rigidement
\`a la notation de Dirac, alors l'expression
sur la droite doit \^etre
r\'e\'ecrite en utilisant l'antisym\'etrie du produit scalaire,
$\langle \Phi | \Psi \rangle =
\langle \Psi | \Phi \rangle ^{\ast}$; ainsi la relation
(\ref{ad}) qui d\'efinit l'adjoint de $A$ devient
\begin{equation}
\langle \Phi | A | \Psi \rangle  =
\langle \Psi | A^{\dag} | \Phi \rangle ^{\ast}
\qquad 
\mbox{pour tout} \ \,
|\Phi \rangle , |\Psi \rangle \in \hs 
\ \ .
\label{adj}
\end{equation}
En cons\'equence,
l'\'el\'ement matriciel
$\langle A | \Phi \rangle , | \Psi \rangle \, \rangle _{\hs}$ peut
seulement \^etre repr\'esent\'e par
$\langle \Psi | A | \Phi \rangle ^{\ast}$ ou par
$\langle \Phi | A^{\dag} | \Psi \rangle$.
Un exemple fr\'equemment utilis\'e est donn\'ee par
\[
\| A | \Psi \rangle \|^2 \; =\;
\langle A | \Psi \rangle , A | \Psi \rangle \rangle _{\hs} \; =\;
\langle \, |\Psi \rangle ,A^{\dag} A|\Psi \rangle \rangle _{\hs}
 \; \equiv \; \langle \Psi |A^{\dag} A |\Psi \rangle
\ \ ,
\]
o\`u la derni\`ere expression est la seule \'ecriture admissible
suivant Dirac.
 
\item
{\em Manque de naturel et de simplicit\'e :}
Comme indiqu\'e dans l'annexe A, on peut se
passer de la discussion de l'espace dual $\hs^{\ast}$
(l'espace des bras),
puisque celui-ci est isom\'etrique \`a $\hs$. Or le
formalisme de Dirac fait un usage syst\'ematique de
$\hs^{\ast}$.
Alors que l'on est habitu\'e \`a des op\'erateurs et matrices
agissant sur ``tout ce qui vient apr\`es", il faut
distinguer dans ce formalisme
entre l'action des op\'erateurs lin\'eaires \`a droite et \`a gauche
\cite{ct},
\begin{eqnarray}
A \left( \lambda | \Phi \rangle +  \mu | \Psi \rangle \right)
& = &
\lambda A | \Phi \rangle +  \mu A | \Psi \rangle
\qquad (\lambda , \mu \in {\bf C} )
\nonumber
\\
\label{aad}
\left( \lambda \langle \Phi | + \mu \langle \Psi | \right) A
& = &
\lambda \langle \Phi | A +
\mu \langle \Psi | A
\ \ ,
\end{eqnarray}
ce qui entra\^{i}ne des ambigu\"\i t\'es potentielles concernant
les domaines de d\'efinition.
Par ailleurs, il faut changer l'ordre naturel des vecteurs dans
certaines expressions qui sont souvent utilis\'ees
(comparer les \'equations (\ref{ad}) et (\ref{adj})).
 
\item
{\em R\`egles de calcul changeantes :}
Quand on passe de $\hs$ \`a $\lt$
(ce que l'on est pratiquement toujours oblig\'e de faire \`a un
certain moment, puisque
la physique a lieu dans l'espace des configurations),
alors une partie des
r\`egles de calcul changent : les op\'erateurs
de diff\'erentiation sur $\lt$ agissent
uniquement vers la droite, leurs \'el\'ements matriciels peuvent
s'\'ecrire comme
$\langle A \varphi , \psi \rangle _{L^2}$, ... etc.
 
\item {\em Interpr\'etation math\'ematique difficile dans un espace
de Hilbert abstrait $\hs$ :}
Si l'on suppose (comme nous l'avons fait) que
les vecteurs appartiennent \`a un espace de Hilbert abstrait
de dimension infinie,
alors on retrouve tous les probl\`emes mentionn\'es dans la
section 3.
Dans ce cas, le formalisme des bras et kets de Dirac repr\'esente
un {\bf calcul purement symbolique}
et ce n'est certainement pas un hasard
que von Neumann
n'a pas cherch\'e une explication ou formulation
math\'ematique de cette approche
en \'elaborant les bases math\'ematiques de
la m\'ecanique quantique \cite{jvn}.
Les introductions modernes \`a ce formalisme
essaient d'en pr\'eciser un peu le contenu math\'ematique,
mais il existe tr\`es peu
de tentatives s\'erieuses
essayant de traduire l'approche
de Dirac dans une th\'eorie math\'ematique rigoureuse suite
\`a une interpr\'etation appropri\'ee de celle-ci
\cite{jr,jpa,gl,jau}.
La th\'eorie r\'esultante
(qui fait intervenir d\`es le d\'epart des triplets
de Gelfand abstraits et des familles spectrales) est tr\`es
compliqu\'ee et difficilement
maniable.
 
\item
{\em Approche non p\'edagogique :}
Les concepts de base de l'alg\`ebre lin\'eaire (applications
lin\'eaires, produits scalaires, ...) sont abondamment
utilis\'es dans tous les domaines de la physique
(m\'ecanique rationnelle, \'electromagn\'etisme, relativit\'e, ...)
avec les notations math\'ematiques standard et non pas avec
le formalisme de Dirac.
L'analyse fonctionnelle bas\'ee sur l'espace
$\lt$ (ou sur $l_2$) est une
synth\`ese naturelle de l'alg\`ebre lin\'eaire et de
l'analyse r\'eelle et des notions de cette th\'eorie
font partie de la formation math\'ematique de tout physicien
(par exemple par le biais de l'analyse de Fourier).
Par contre le calcul symbolique de Dirac
donne parfois l'impression de repr\'esenter quelque chose
de qualitativement nouveau et pratiquement
incontournable pour le d\'eveloppement de la m\'ecanique
quantique\footnote{Il est peut-\^etre bon de rappeler
que la th\'eorie quantique a \'et\'e
d\'evelopp\'ee sans l'utilisation de ce formalisme \cite{vdw}
et de noter que
son enseignement peut s'en passer largement ou
compl\`etement comme en t\'emoigne un bon nombre
d'excellents
ouvrages \cite{gap,ll,pd}.}.
\end{itemize}
 
Pour les espaces de Hilbert de dimension
finie qui interviennent dans la des\-cription du spin (ou du moment 
angulaire) des particules,
on a $\hs \simeq {\bf C}^n$ et la notation de Dirac
repr\'esente alors une r\'e\'ecriture des vecteurs et une
interpr\'etation
particuli\`ere des op\'erations
de l'alg\`ebre
lin\'eaire standard \cite{fano}.
Dans ce cas, tout est math\'ematiquement bien d\'efini, mais
les autres inconv\'enients mentionn\'es ne sont pas tous \'ecart\'es.

\subsection{Avantages}
 
La grande puissance des notations de Dirac consiste dans le fait
qu'elles permettent de faire des calculs formels donnant automatiquement
la forme correcte des r\'esultats.
Par exemple, l'insertion de l'op\'erateur identit\'e (\ref{id}) entre
deux op\'erateurs lin\'eaires $A$ et $B$,
\begin{equation}
\label{in1}
\langle \Phi | AB| \Psi \rangle 
\stackrel{(\ref{id})}{=}
\langle \Phi | A
\left( \sum_{n=1}^{\infty} |n \rangle
\langle n | \right)  B| \Psi \rangle
= \sum_{n=1}^{\infty}
\langle \Phi | A| n \rangle
\langle n | B| \Psi \rangle
\ \ ,
\end{equation}
donne tout de suite le bon r\'esultat final sans qu'on ait besoin
de contempler l'action successive des applications
$|n \rangle$ et $\langle n |$ d\'ecrite dans l'\'equation
(\ref{de}). 
Simi\-lairement, le projecteur $P_{|\Phi \rangle}$
sur l'\'etat $|\Phi \rangle \in \hs$ s'\'ecrit simplement
\begin{equation}
\label{in2}
P_{|\Phi \rangle} =
|\Phi \rangle \langle \Phi |
\]
et pour ses \'el\'ements de matrice on obtient imm\'ediatement
\[
\langle \Psi | P_{|\Phi \rangle} | \Psi^{\prime} \rangle =
\langle \Psi | \Phi \rangle  \langle \Phi | \Psi^{\prime} \rangle
\ \ .
\end{equation}

\subsection{``Solution" des probl\`emes}
 
Comme nous venons de le souligner,
la notation $| \Psi \rangle$ pour les vecteurs
et $\langle \Psi |$ pour les formes lin\'eaires est tr\`es
utile \`a des fins mn\'emotechniques et calculatoires.
Aussi serait-il
d\'eplac\'e d'\'eviter ces notations
et de se priver de leurs
avantages.
Un bon compromis que nous
r\'esumerons maintenant
est celui adopt\'e ou mentionn\'e
dans un certain nombre d'ouvrages \cite{gap,ll,pd,th}.
 
Si nous ``identifions"
les diff\'erents espaces de Hilbert discut\'es dans la section 2,
nous pouvons \'ecrire 
$\hs = \lt$ et dans ce cas nous \'evitons
d\'ej\`a les complications math\'ematiques
du formalisme invariant (section~3).
Dans tous les cas
- que l'on fasse cette identification ou pas - il est
souvent commode
de {\em noter les fonctions d'onde par} $| \psi \rangle$
au lieu de $\psi$
(ou par $\, \psi \rangle$ comme sugg\'er\'e par Dirac \cite{d})
pour m\'emoriser les relations suivantes
qui sont valables pour toute base orthonorm\'ee
$\{ | \varphi_n \rangle \}_{n\in {\bf N}}$ de $\lt$ :
\begin{eqnarray*}
\langle \, |\varphi_n \rangle \, ,\, |\varphi_m \rangle \, \rangle _{L^2}
\equiv
\langle \varphi_n  | \varphi_m \rangle & = & \delta_{nm}
\\
\sum_{n \in {\bf N}}
| \varphi_n \rangle \langle \varphi_n | & = & {\bf 1}_{L^2}
\ \ .
\end{eqnarray*}
Ici, la derni\`ere relation veut dire que
\begin{eqnarray*}
|\psi \rangle & = &
\sum_{n \in {\bf N}}
| \varphi_n \rangle \langle \varphi_n | \psi \rangle
\qquad  \ \mbox{pour tout} \ \  | \psi \rangle \in \lt
\\
\mbox{ou} \quad
| \psi (x) \rangle & = &
\sum_{n \in {\bf N}}
| \varphi_n (x) \rangle \langle \varphi_n | \psi \rangle
\quad \mbox{pour tout} \ \  x \in \br
\ \ .
\end{eqnarray*}
Dans le m\^eme esprit, le projecteur $P_{\psi}$ sur
$| \psi \rangle \in \lt$ peut s'\'ecrire comme
$P_{\psi} = | \psi \rangle \langle \psi |$.
 
Pour les op\'erateurs, il est commode d'utiliser la notation \cite{ct}
\[
| A \psi \rangle \equiv  A | \psi \rangle
\]
tout en \'evitant l'interpr\'etation (\ref{cohe}) 
des \'el\'ements matriciels qui est source 
d'ambigu\"{i}t\'es;  
un \'el\'ement matriciel
peut alors \^etre mis sous une des formes suivantes :
\[
\langle \varphi | A | \psi \rangle  =
\langle \varphi | A  \psi \rangle  =
\langle A^{\dag} \varphi | \psi \rangle =
\langle \psi | A^{\dag} \varphi \rangle ^{\ast} =
\langle \psi | A^{\dag} | \varphi \rangle ^{\ast}
\ \ .
\]
Les insertions d'op\'erateurs sont r\'ealis\'ees comme
dans les expressions
(\ref{in1}) et (\ref{in2}).
 
Finalement, la
notation $|n..m \rangle$ au lieu de $\varphi_{n..m}$ pour
les vecteurs d'une base de l'espace de Hilbert,  
index\'ee par $n,..,m$,
est bien utile pour \'ecrire les \'el\'ements 
matriciels\footnote{Il faut prendre garde au fait que la matrice 
r\'esultante repr\'esente un op\'erateur lin\'eaire 
sur l'espace de Hilbert $l_2$ qui est de dimension infinie, ce qui 
implique qu'il faut se soucier de son domaine de d\'efinition : 
l'existence de repr\'esentations matricielles et leurs pi\`eges 
math\'ematiques sont discut\'es dans la r\'ef\'erence \cite{sg}.},
\[
a_{n..m,n^{\prime} .. m^{\prime}} =
\langle n..m | A |
n^{\prime} .. m^{\prime} \rangle
\ \ .
\]
Ainsi, avec un peu de flexibilit\'e,
on peut b\'en\'eficier des avantages de la notation
de Dirac tout en \'evitant ses inconv\'enients.
 
\section{Conclusion}
 
Essayons de tirer quelques conclusions des discussions
pr\'ec\'edentes, en particulier
pour l'enseignement de la m\'ecanique quantique.
 
Physique et math\'ematique sont deux sciences diff\'erentes
et on peut tr\`es bien
justifier qu'une pr\'esentation de physicien ne tienne pas compte
d'une rigueur math\'ematique parfaite m\^eme si l'auteur
ma\^itrise compl\`etement celle-ci. En physique
c'est probablement un art d'utiliser un minimum de math\'ematiques
tout en restant tellement pr\'ecis dans le raisonnement et
dans la pr\'esentation que
le physicien math\'ematicien puisse compl\'eter tous les d\'etails
techniques
sans ambigu\"\i t\'e et ainsi \'etablir les r\'esultats et leur
domaine de validit\'e d'une mani\`ere irr\'efutable.
En m\'ecanique quantique, une telle d\'emarche
correspondrait \`a donner des d\'efinitions pr\'ecises au
d\'epart (pour les op\'erateurs lin\'eaires sur $\lt$)
tout en
\'evitant de discuter syst\'ematiquement les d\'etails math\'ematiques
(domaines de d\'efinition, distributions, ...) dans la suite.
Par contre,
toute approche bas\'ee sur un calcul symbolique qui est
tr\`es difficile \`a rendre rigoureux (et donc apte \`a des
conclusions pr\'ecises) para\^it contestable.
Ceci est d'autant plus vrai que la premi\`ere approche
n'est pas plus compliqu\'ee et qu'elle est bas\'ee sur
une th\'eorie math\'ematique
standard, bien d\'evelopp\'ee et trouvant des applications dans
beaucoup d'autres domaines de la physique (syst\`emes dynamiques,
relativit\'e, optique, ...).
 
Les ouvrages de physique
qui ne suivent pas l'approche du calcul
symbo\-lique (et qui mentionnent les domaines de d\'efinition
ainsi que la diff\'erence entre op\'erateurs hermitiens
et auto-adjoints) ne sont pas tr\`es nombreux : citons
les monographies \cite{gap} qui ne sont pas bas\'ees
sur le formalisme invariant et qui font un usage
lib\'eral des notations de Dirac chaque fois que cela para\^it
utile (voir aussi \cite{bal}).
Une pr\'esentation comparable,
mais plus math\'ematique et s'orientant vers les fondements
conceptuels est donn\'ee dans
\cite{jmj}, alors que les trait\'es
\cite{rj,ber,th,gl}
peuvent \^etre qualifi\'es de relevant du domaine de la
physique math\'ematique.
Parmi les ouvrages \cite{ct}, ceux de Messiah, Peebles et Schwabl
discutent
en d\'etail la m\'ecanique ondulatoire et sa structure
g\'eom\'etrique avant de pr\'esenter le formalisme invariant
et les notations de Dirac.
Finalement nous mentionnons aussi quelques ouvrages
qui \'evitent le formalisme
invariant et l'usage rigide des
notations de Dirac, mais qui
ne discutent pas les d\'etails math\'ematiques concernant les 
op\'erateurs sur $\lt$ : \`a part les `classiques' \cite{ll},
il s'agit des introductions \'el\'ementaires et modernes
\cite{pd} qui pr\'esentent clairement tous les principes
de la th\'eorie tout en appliquant
un strict minimum de
math\'ematiques utiles.

\bigskip 
\bigskip 
 
%\vspace{14mm}

\newpage 

{\bf \Large Remerciements}
 
%\vspace{5mm}
 
\bigskip 

Je profite de l'occasion pour remercier les
enseignants superbes aupr\`es de qui j'ai pu apprendre la m\'ecanique
quantique \`a G\"ottingen (H.Goenner, H.Roos, F.Hund ($\dagger$1997),
J.Yngvason, H.Reeh) et \`a Berkeley (G.W.Mackey).
Merci \`a A.Bilal pour m'avoir expliqu\'e un peu le formalisme
invariant lors de mon arriv\'ee en France!
J'exprime ma gratitude \`a I.Laktineh, M.Kibler
et R.Barbier pour avoir endur\'e mes innombrables complaintes,
pour toutes nos discussions sur la m\'ecanique quantique et pour
leurs
commentaires sur le texte. Merci \`a D.Maison, P.Breitenlohner, 
S.Fleck, R.McDermott,
H.K\"uhn et au tr\`es regrett\'e K.Baumann ($\dagger$1998) 
pour leurs remarques pertinentes sur les exemples!
Je tiens aussi \`a exprimer 
ma gratitude \`a
B\'en\'edicte Bruckert et Annabelle
Pontvianne pour de nombreuses discussions sur
la physique et les math\'ematiques.

\newpage
 
\appendix
 
\section{Le formalisme math\'ematique}
 
\subsection{Les notations de Dirac}
 
Un vecteur $|\Psi \rangle \in \hs$ est appel\'e {\em ket}$\,$
et \`a ce vecteur nous pouvons associer une forme lin\'eaire
$\omega_{|\Psi \rangle} \equiv \langle \Psi |$
appel\'ee {\em bra}$\,$ et d\'efinie par l'interm\'ediaire
du produit scalaire ($\!${\em ``bracket"}$\,$) :
\begin{eqnarray}
\label{bra}
\omega_{|\Psi \rangle} \equiv \langle \Psi |
\ : & \hs & \stackrel{{\rm lin.}}{\longrightarrow} \ \ {\bf C}
\\
 & |\Phi \rangle & \longmapsto \ \
\omega_{|\Psi \rangle} \left( | \Phi \rangle \right)
= \langle \,
|\Psi \rangle \, , \,   | \Phi \rangle \, \rangle  _{\hs} \equiv
\langle \Psi | \Phi \rangle
\ \ .
\nonumber
\end{eqnarray}
D'apr\`es l'in\'egalit\'e de Cauchy et Schwarz reliant le produit
scalaire et la norme
$\|  \Psi  \| \equiv \| |  \Psi  \rangle \| =
\sqrt{\langle \Psi | \Psi \rangle}$
dans $\hs$,
\[
| \langle \Psi | \Phi \rangle | \leq
\| \Psi \|   \cdot
\| \Phi  \|
\ \ ,
\]
la forme
lin\'eaire $\omega_{|\Psi \rangle}$ est continue : ceci veut dire que
pour tout $|\Phi \rangle \in \hs$ il existe une constante
$c\geq 0$ telle que
$| \omega_{| \Psi \rangle} (| \Phi \rangle ) | \leq
c  \, \| \Phi \|$. Par cons\'equent, le bra
$\langle \Psi |$ est un \'el\'ement de l'{\em espace
de Hilbert dual}
\[
\hs^{\ast} = \{ \omega : \hs
\longrightarrow {\bf C} \ \; \mbox{lin\'eaire et continu} \}
\ \ .
\]
 
Inversement, \`a tout bra $\langle \Psi | \in \hs^{\ast}$
nous pouvons associer un ket $| \Psi \rangle \in \hs$;
en effet, d'apr\`es le lemme de Riesz \cite{rs},
chaque \'el\'ement $\omega \in \hs^{\ast}$ d\'etermine
de mani\`ere unique un vecteur $| \Psi _{\omega} \rangle
\in \hs$ tel que
\[
\omega \left( | \Phi \rangle \right)
= \langle \Psi_{\omega} | \Phi \rangle
\qquad \mbox{pour tout} \quad | \Phi \rangle \in \hs
\ \ .
\]
(Le vecteur
$| \Psi _{\omega} \rangle$ ``r\'ealise" l'application $\omega$
par l'interm\'ediaire du produit scalaire.)
Le vecteur associ\'e \`a la forme lin\'eaire $\langle \Psi |$
est not\'e par $| \Psi \rangle$ et nous avons donc
une correspondance biunivoque entre $| \Psi \rangle \in \hs$
et $\langle \Psi | \in \hs^{\ast}$ :
\begin{equation}
\fbox{\mbox{$ \ \hs \ni
| \Psi \rangle \ \ \stackrel{{\rm 1-1}}{\longleftrightarrow}
\ \ \langle \Psi | \in \hs^{\ast} \ $}}
\ \ .
\end{equation}
Ainsi nous pouvons identifier\footnote{Si on d\'efinit la norme
de $\omega \in \hs^{\ast}$ par $\| \omega \| =
{\rm sup} \, | \omega (f) |$ (o\`u le supremum est pris
sur tous les vecteurs 
$f \in \hs$ de norme $1$), alors
on peut montrer que la bijection $\hs \to \hs^{\ast}$
est antilin\'eaire et qu'elle
pr\'eserve les normes, c'est-\`a-dire qu'elle repr\'esente
une isom\'etrie.} $\hs$ et $\hs^{\ast}$ et nous passer
compl\`etement de $\hs^{\ast}$. L'introduction d'un espace vectoriel
dual est seulement n\'ecessaire pour d\'efinir les vecteurs
g\'en\'eralis\'es, voir section A.3 ci-dessous. 
 
Une {\em base hilbertienne}
$\left\{ | \Phi_n \rangle \equiv |n\rangle \right\}_{n\in \bn}$
de $\hs$ est un ensemble de vecteurs satis\-faisant
la relation d'orthonormalit\'e
\begin{equation}
\langle n | m \rangle = \delta_{nm}
\qquad \mbox{pour tout} \quad n,m \in \bn
\end{equation}
et la relation de fermeture
\begin{equation}
\label{id}
\sum_{n=0}^{\infty} | n \rangle \langle n | =  {\bf 1}_{\hs}
\ \ .
\end{equation}
Cette relation fait intervenir la somme des op\'erateurs
$| n \rangle \langle n |$ qui sont obtenus par composition de
deux applications :
\begin{eqnarray}
\label{de}
\hs \ \ \stackrel{\langle n|}{\longrightarrow} & {\bf C}&
\stackrel{|n \rangle}{\longrightarrow} \quad \ \ \hs
\\
| \Phi \rangle \ \longmapsto & \langle n | \Phi \rangle &
\longmapsto \
\langle n | \Phi \rangle \,  |n \rangle
\ \ .
\nonumber
\end{eqnarray}
Ici, la premi\`ere application est la forme lin\'eaire
(\ref{bra}) et la deuxi\`eme repr\'esente la multiplication d'un
nombre complexe par le vecteur $|n\rangle \in \hs$.
 
Un {\em op\'erateur sur $\hs$} est une application
lin\'eaire
\begin{eqnarray}
A \ : & {\cal D} (A) & \longrightarrow \ \ \hs
\\
 & |\Psi \rangle & \longmapsto \ \
A | \Psi \rangle
\ \ ,
\nonumber
\end{eqnarray}
o\`u ${\cal D}(A)$ ({\em domaine de d\'efinition
de} $A$) est un sous-espace vectoriel dense
de $\hs$.
(On peut g\'en\'eraliser cette d\'efinition en supprimant
les hypoth\`eses que ${\cal D} (A)$ soit dense et que
l'application $A$ soit lin\'eaire, mais ces g\'en\'eralisations
n'interviennent gu\`ere en m\'ecanique quantique.)
 
Avant de discuter quelques exemples d'op\'erateurs, nous
rappelons que
le produit scalaire des vecteurs $|\Phi \rangle$ et $A | \Psi
\rangle$ est not\'e suivant Dirac par
\begin{equation}
\label{mat}
\langle \, | \Phi \rangle \, ,\, A|\Psi \rangle \, \rangle _{\hs} \equiv
\langle \Phi | A | \Psi \rangle
\ \ .
\end{equation}
L'expression $\langle \Phi | A | \Psi \rangle$
peut donc \^etre consid\'er\'e comme r\'esultant de la composition
de deux applications lin\'eaires,
\begin{eqnarray}
\label{copo}
{\cal D} (A) \ \ \stackrel{A}{\longrightarrow} & \hs &
\stackrel{\langle \Phi |}{\longrightarrow} \quad \ \ {\bf C}
\\
| \Psi \rangle \ \longmapsto & A | \Psi \rangle &
\longmapsto \
\langle \Phi | A | \Psi \rangle
\ \ ,
\nonumber
\end{eqnarray}
o\`u la composition est d\'efinie comme d'habitude par
$\left( \langle \Phi | \circ A \right)  | \Psi \rangle :=
\langle \Phi | \left(  A   | \Psi \rangle \right)$.
Cependant Dirac ne s'est pas limit\'e \`a 
cette interpr\'etation sans \'equivoque des notations
qu'il a introduites 
- voir section 4.1.

\subsection{Op\'erateurs lin\'eaires}
 
Pour simplifier l'\'ecriture et \'eviter toute ambigu\"\i t\'e,
nous n'utilisons pas les notations de Dirac dans la suite.
Nous encourageons fortement le lecteur qui n'est pas familier avec les
d\'efinitions et r\'esultats \'enonc\'es au d\'ebut de cette section 
de continuer la lecture avec les nombreuses 
illustrations qui suivront.

Pour un op\'erateur $A$ sur $\hs$,
le {\em domaine de d\'efinition de} $A^{\dag}$
est d\'efini par
\begin{eqnarray*}
{\cal D} (A^{\dag}) & = & \{ \varphi \in \hs \, | \, \exists\,
\tilde{\varphi}(A;\varphi ) \in \hs \ \ \mbox{tel que}
\\
&& \qquad \qquad \qquad
\left\langle \varphi , A \psi \right\rangle =
\langle \, \tilde \varphi (A;\varphi ) \, , \, \psi \, \rangle
\ \mbox{pour tout} \ \psi  \in {\cal D} (A) \}
\ \ .
\end{eqnarray*}
(La notation
$\tilde \varphi (A;\varphi )$ indique que le vecteur $\tilde{\varphi}$
d\'epend de $A$ et de $\varphi$.)
Pour $\varphi \in {\cal D}(A^{\dag})$, on d\'efinit
$A^{\dag} \varphi = \tilde \varphi (A;\varphi )$, c'est-\`a-dire
\begin{equation}
\label{defa}
\langle \varphi , A \psi \rangle =
\langle A^{\dag} \varphi , \psi \rangle
\qquad \mbox{pour tout} \ \psi \in {\cal D}(A)
\ \ .
\end{equation}
 
En th\'eorie quantique,
les observables physiques sont d\'ecrites par des op\'erateurs
$A$ sur $\hs$ qui ont la propri\'et\'e d'\^etre
{\em auto-adjoint} : ceci veut dire que
$A=A^{\dag}$, c'est-\`a-dire ${\cal D} (A) = {\cal D} (A^{\dag})$
et $A\varphi =A^{\dag} \varphi$ pour tout $\varphi \in {\cal D} (A)$.
Cette condition assure que le
spectre de  $A$ est r\'eel
et que les vecteurs propres (g\'en\'eralis\'es)
de $A$ forment un syst\`eme complet de vecteurs orthonormaux.

Le {\em spectre} d'un op\'erateur auto-adjoint est l'union du 
{\em spectre discret} (ensemble des valeurs propres de $A$) et du  
{\em spectre continu} (ensemble des valeurs propres g\'en\'eralis\'ees de
$A$, c'est-\`a-dire des valeurs propres pour lesquelles les 
vecteurs propres n'appartiennent pas
\`a $\hs$) : ces notions seront pr\'ecis\'ees et illustr\'ees 
dans la suite (ainsi que dans l'annexe C o\`u nous mentionnerons aussi 
le soi-disant spectre r\'esiduel qui peut appara\^{i}tre pour un  
op\'erateur non auto-adjoint).

Deux complications
techniques apparaissent lors de l'\'etude d'une observable $A$
en m\'ecanique quantique :
 
\noindent (i) Si le spectre de $A$ n'est pas born\'e, alors
le domaine de d\'efinition de $A$ ne peut pas \^etre $\hs$
tout entier.
 
\noindent (ii) Si le spectre de $A$ contient une partie continue, 
alors les vecteurs propres correspondants
n'appartiennent pas \`a $\hs$,
 mais \`a un espace plus large contenant $\hs$.
 
Dans cette section et dans la suivante,
nous discuterons tour \`a tour ces deux probl\`emes.
 
La classe la plus simple d'op\'erateurs est celle
des op\'erateurs qui sont {\em born\'es}, c'est-\`a-dire
pour tout vecteur $\psi \in {\cal D}(A)$, on a
\begin{equation}
\label{bor}
\| A  \psi \| \leq c \, \| \psi \|
\qquad \mbox{o\`u $c\geq 0$ est une constante}
\ \ .
\end{equation}
Cette condition revient \`a dire
que le spectre de $A$ est born\'e. Les op\'erateurs born\'es
peuvent toujours \^etre d\'efinis sur tout l'espace
de Hilbert, c'est-\`a-dire que
${\cal D}(A)= \hs$. Un exemple important est
celui d'un op\'erateur unitaire $U : \hs \to \hs$ ; un tel
op\'erateur est born\'e, car
la relation (\ref{un}) implique
$\| U  \psi \| = \| \psi \|$
pour tout $\psi \in \hs$ et la condition (\ref{bor})
est donc satisfaite. (Le spectre de $U$ est born\'e, puisqu'il 
appartient au cercle unit\'e du plan complexe.) 
 
Une grande partie
des subtilit\'es math\'ematiques de la m\'ecanique quantique
provient du r\'esultat suivant \cite{sg, rs}.
\begin{theo}[Hellinger et Toeplitz]
Soit $A$ un op\'erateur sur $\hs$ qui est partout d\'efini
et qui satisfait \`a la condition d'hermicit\'e
\begin{equation}
\label{her}
\langle \varphi , A  \psi \rangle =
\langle A  \varphi  ,   \psi \rangle
\end{equation}
pour tous les vecteurs $\varphi, \psi \in \hs$.
Alors $A$ est born\'e.
\end{theo}
En th\'eorie quantique, on a souvent des op\'erateurs
comme ceux de position, d'impulsion ou d'\'energie qui
ob\'eissent
\`a la condition d'hermicit\'e
(\ref{her})
sur leur domaine de d\'efinition,
mais pour lesquels
le spectre n'est pas born\'e. 
(En fait, la relation structurelle de base de la m\'ecanique quantique,
c'est-\`a-dire la relation de commutation canonique, 
exige m\^eme que certains des op\'erateurs fondamentaux  
qu'elle fait intervenir  soient non born\'es - voir 
annexe C.)
Le th\'eor\`eme pr\'ec\'edent
indique alors qu'il n'est pas possible de d\'efinir ces
op\'erateurs hermitiens
sur tout l'espace de Hilbert $\hs$ et que leur
domaine de d\'efinition doit n\'ecessairement \^etre
un vrai sous-espace de $\hs$. Parmi tous les
choix de sous-espace qui sont math\'ematiquement possibles,
certains sont
privil\'egi\'es en pratique par des consid\'erations
physiques (conditions aux limites, ...)
\cite{aw,rs,sg,ber,th}.
 
A titre d'exemple,
consid\'erons {\em l'op\'erateur de position} $Q$, c'est-\`a-dire 
l'op\'erateur `multiplication par $x$' sur
l'espace de Hilbert $\lt$ :
\begin{equation}
\label{pos}
\left(Q \psi \right) (x) = x \, \psi (x)
\quad  \mbox{pour tout $x\in \br$}
\ \ .
\end{equation}
Le {\em domaine de d\'efinition maximal} de $Q$ est
celui qui assure que la fonction $Q\psi$ existe et 
qu'elle appartient encore \`a
$\lt$ :
\begin{equation}
\label{dq}
{\cal D}_{\rm max} (Q) = \{ \psi \in \lt \ | \
\| x \psi \|^2 \equiv \int_{\br}dx \, x^2 | \psi (x) |^2
< \infty \}
\ \ .
\end{equation}
Pour tous les vecteurs de cet espace (qui est un sous-espace non-trivial 
et dense de
$\lt$), la condition (\ref{her}) est satisfaite ce qui implique
que le spectre de $Q$ est r\'eel. En fait, le spectre
de cet op\'erateur est
constitu\'e de tout l'axe r\'eel et il n'est donc pas born\'e.
 
Nous notons que
pour certaines consid\'erations il est commode de disposer
d'un domaine de d\'efinition qui est laiss\'e invariant
par l'op\'erateur. Pour l'op\'erateur $Q$, un tel domaine
est donn\'e par
l'espace de Schwartz $\st$
des fonctions \`a d\'ecroissance rapide.
Rappelons qu'une
fonction $f : \br \to \bc$ appartient \`a $\st$ si elle
est d\'erivable une infinit\'e de fois et si cette fonction et toutes
ses d\'eriv\'ees d\'ecroissent plus vite \`a l'infini
que l'inverse d'un polyn\^ome quelconque).  On peut en d\'eduire que 
$\st \subset {\cal D}_{\rm max} (Q)$ et
\[
Q \ : \ \st \longrightarrow \st
\ \ .
\]
L'espace de Schwartz
est aussi un {\em domaine de d\'efinition invariant} pour 
{\em l'op\'erateur
d'impulsion} $P = \ds{\hbar \over \ri} \, \ds{d \over dx}$
sur $\lt$, c'est-\`a-dire
$P \, : \, \st \rightarrow \st$.

\subsection{Triplets de Gelfand (vecteurs g\'en\'eralis\'es)}
 
L'op\'erateur de position
(\ref{pos}) d\'efini sur $\st$ illustre aussi
le fait que les vecteurs propres associ\'es au spectre
continu d'un op\'erateur
auto-adjoint n'appartiennent pas \`a l'espace
de Hilbert\footnote{Pour \^etre pr\'ecis, l'op\'erateur (\ref{pos})
d\'efini sur $\st$ est {\it essentiellement auto-adjoint},
ce qui implique qu'on peut le rendre auto-adjoint d'une seule
fa\c con en \'elargissant
son domaine de d\'efinition de mani\`ere naturelle
(voir \cite{rs, sg}
pour des d\'etails).}.
En effet,
la fonction propre $\psi_{x_0}$ associ\'ee \`a la valeur
propre $x_0 \in \br$ est d\'efinie par la
relation
\begin{equation}
\label{vpq}
\left( Q \psi_{x_0} \right) (x) = x_0 \, \psi_{x_0} (x)
\qquad (x_0 \in \br \ ,  \ \psi_{x_0} \in \st \ , \
\psi_{x_0} \not \equiv 0 )
\end{equation}
ou encore, d'apr\`es (\ref{pos}), par
\[
(x-x_0) \, \psi_{x_0} (x) = 0
\quad  \mbox{pour tout $x\in \br$}
\ \ .
\]
Cette condition implique
$\psi_{x_0} (x) =0$ pour $x \neq x_0$. En cons\'equence, la fonction
$\psi_{x_0}$ est nulle presque partout et repr\'esente donc l'\'el\'ement
nul de $\lt$ \cite{sg, af, rs}.
L'op\'erateur $Q$ n'admet donc aucune valeur propre.
 
Notons que la situation est la m\^eme pour l'op\'erateur
$P$ d\'efini sur $\st$ qui est aussi
essentiellement auto-adjoint :
l'\'equation aux valeurs propres
\[
\left( P\psi_p \right) (x) = p \, \psi_p (x)
\qquad ( p \in \br \ , \ \psi_p \in \st \ , \
\psi_p \not \equiv 0 ),
\ \ ,
\]
est r\'esolue par $\psi_p (x) = 1/\sqrt{2\pi \hbar}
\  {\rm exp} (\ri px / \hbar )$,
mais $\psi_p \not \in \st$.
Donc $P$ n'admet aucune valeur propre.
 
Par contre, les \'equations aux valeurs propres pour $Q$ et $P$
admettent des solutions faibles (solutions distributives).
Par exemple, la
fonction g\'en\'eralis\'ee (distribution) de Dirac avec support en $x_0$,
$\delta_{x_0} (x) \equiv \delta (x-x_0)$, est une solution faible
de l'\'equation aux valeurs propres (\ref{vpq}) :
pour v\'erifier que $x \, \delta _{x_0} (x) = x_0 \, \delta_{x_0} (x)$
au sens des distributions, il faut \'etaler cette relation avec une
fonction test $\varphi \in \st$ :
\begin{equation}
\label{for}
\int_{\br} dx \,
x \, \delta _{x_0} (x) \, \varphi (x) = x_0 \, \varphi (x_0) =
\int_{\br} dx \,
x_0 \, \delta _{x_0} (x) \, \varphi (x)
\ \ .
\end{equation}
La fonction g\'en\'eralis\'ee de Dirac et la fonction g\'en\'eralis\'ee
$x \delta _{x_0}$ n'appartiennent pas
au domaine de d\'efinition $\st$ de $Q$,
mais \`a l'espace dual
\[
\std = \{ \omega : \st \to {\bf C} \
\mbox{lin\'eaire et continu} \}
\ \ ,
\]
c'est-\`a-dire \`a l'espace des distributions temp\'er\'ees
sur $\br$ \cite{rs, sg, bgc, ls, gv}.
Elles sont d\'efinies d'une mani\`ere abstraite et pr\'ecise par
\begin{eqnarray}
\label{xd}
\delta_{x_0} \ : & \st  & \longrightarrow \ \ \bc
\\
 & \varphi & \longmapsto \ \
\delta_{x_0} (\varphi ) = \varphi (x_0)
\nonumber
\end{eqnarray}
et
\begin{eqnarray*}
x\, \delta_{x_0} \ : & \st  & \longrightarrow \ \ \bc
\\
 & \varphi & \longmapsto \ \
\left( x\, \delta_{x_0} \right) (\varphi ) = \delta_{x_0} (x \varphi )
\stackrel{(\ref{xd})}{=} x_0 \, \varphi (x_0)
\ \ .
\end{eqnarray*}
Avec ces d\'efinitions,
l'\'ecriture formelle (\ref{for}) prend la forme exacte
\[
\left(
x \, \delta _{x_0} \right) (\varphi) = \left(
x_0 \, \delta _{x_0} \right) (\varphi)
\qquad \mbox{pour tout} \ \ \varphi \in \st
\ \ .
\]
Ainsi
l'\'equation aux valeurs propres $Q \psi_{x_0} = x_0 \, \psi_{x_0}$
admet une solution distributive
$\psi_{x_0}$ pour tout nombre $x_0 \in \br$.
Comme le spectre de l'op\'erateur (essentiellement
auto-adjoint) $Q$ est l'ensemble
des nombres r\'eels pour lesquels l'\'equation aux valeurs propres
admet comme solution une fonction
$\psi \in {\cal D}(Q) = \st$ (spectre discret)
ou bien une fonction
g\'en\'eralis\'ee $\psi \in \std$ (spectre continu),
nous pouvons conclure que ${\rm Sp} \, Q = \br$ et que 
le spectre de $Q$ est
purement continu.
 
De mani\`ere analogue, la fonction
$\psi_p (x) = 1/\sqrt{2\pi \hbar}
\  {\rm exp} ( \ri px / \hbar)$
d\'efinit une distribution $l_p$ selon
\begin{eqnarray}
\label{dis}
l_p  \ : & \st  & \longrightarrow \ \ \bc
\\
 & \varphi & \longmapsto \ \
l_p (\varphi ) = \int_{\br} dx \, \overline{\psi_p (x)} \,
\varphi (x) 
\stackrel{(\ref{f})}{=} ({\cal F} \varphi ) (p)
\ \ ,
\nonumber
\end{eqnarray}
o\`u ${\cal F} \varphi$ d\'enote la transform\'ee
de Fourier (\ref{f}).
 La distribution $l_p$ repr\'esente une solution de l'\'equation
aux valeurs propres $Pl_p = p \, l_p$, car d'apr\`es 
les r\`egles de calcul pour les distributions et la transformation 
de Fourier \cite{sg, gv} et d'apr\`es la d\'efinition (\ref{dis}), nous
avons \[
\left( Pl_p \right) (\varphi) 
=
\left( \ds{\hbar \over \ri} \, \ds{dl_p \over dx} \right) (\varphi)
=
l_p \left( \ds{\hbar \over \ri} \, \ds{d\varphi \over dx} \right) 
\stackrel{(\ref{dis})}{=}
({\cal F}
\left( \ds{\hbar \over \ri} \, \ds{d\varphi \over dx} \right) ) (p)
= 
p \, ( {\cal F} \varphi ) (p) 
\stackrel{(\ref{dis})}{=} 
p \, l_p (\varphi).
\]
Il s'ensuit que ${\rm Sp} \, P = \br$ (spectre purement continu).
 
Le probl\`eme aux valeurs propres pour des op\'erateurs
avec spectre continu nous am\`ene
donc \`a consid\'erer
le {\em triplet de Gelfand}
({\em ``rigged Hilbert space"} ou {\em triade
hilbertienne})\footnote{Les triplets de Gelfand
sont discut\'es en d\'etail dans l'ouvrage \cite{gv}
(voir aussi \cite{jr} pour une d\'efinition l\'eg\`erement
modifi\'ee). Une
introduction courte et excellente aux d\'efinitions
et applications en m\'ecanique quantique est donn\'ee dans
les r\'ef\'erences
\cite{blt,sg,ber}. Concernant l'importance
des triplets de Gelfand, nous citons
leurs inventeurs \cite{gv} : ``Nous estimons que cette
notion est, au moins, aussi importante, sinon plus,
que la notion d'espace de Hilbert."}
\begin{equation}
\label{tri}
\fbox{\mbox{$ \
\st \subset \lt \subset \std \ $}}
\ \ .
\end{equation}
Ici, $\st$ est un sous-espace dense de $\lt$ \cite{rs}
et toute fonction
$\psi \in \lt$ d\'efinit une distribution $\omega_{\psi} \in \std$
selon
\begin{eqnarray}
\omega_{\psi} \ : & \st & \longrightarrow \ \ {\bf C}
\nonumber
\\
 & \varphi & \longmapsto \ \
\omega_{\psi} (\varphi) = \int_{\br} dx \, \overline{\psi(x)}
\varphi (x)
\ \ .
\label{reg}
\end{eqnarray}
Mais $\std$ contient aussi des distributions
comme la distribution $\delta_{x_0}$ de Dirac ou la distribution $l_p$
qui ne peuvent pas \^etre
repr\'esent\'ees \`a l'aide d'une fonction $\psi \in \lt$ selon
(\ref{reg}).
La proc\'edure d'\'etalement avec une fonction test $\varphi \in \st$
correspond
\`a la formation de {\em paquets d'onde} et la th\'eorie des
distributions donne un sens bien pr\'ecis \`a cette proc\'edure
ainsi qu'aux fonctions g\'en\'eralis\'ees qu'elle fait intervenir.
 
La d\'efinition abstraite
du triplet (\ref{tri}) peut \^etre pr\'ecis\'ee (topologie
de $\st$,...) et par ailleurs
$\st$ peut \^etre g\'en\'eralis\'e \`a d'autres
sous-espaces (associ\'es avec $Q$ ou avec d'autres
op\'erateurs d\'efinis sur $\lt$).
A ces d\'etails pr\`es, nous pouvons dire :
\begin{quote}
{\bf Le triplet (\ref{tri})
d\'ecrit d'une mani\`ere exacte et simple la nature
math\'ematique de tous les kets et bras utilis\'es en m\'ecanique
quantique.}
\end{quote}
En effet, d'apr\`es le lemme de Riesz mentionn\'e dans la section A.1,
l'espace de Hilbert $\lt$ est \'equivalent \`a son dual : \`a
tout ket appartenant \`a $\lt$
correspond donc un bra et inversement. Par ailleurs,
un ket appartenant au sous-espace $\st$ d\'efinit toujours un bra
appartenant \`a $\std$ selon la d\'efinition (\ref{reg}). Mais
il existe des \'el\'ements de $\std$, des {\em bras g\'en\'eralis\'es},
auxquels on ne peut pas associer un ket appartenant \`a
$\st$ ou $\lt$. Nous notons que
la transparence de ce r\'esultat math\'ematique se perd,
si l'on proc\`ede comme d'habitude et que l'on \'ecrit l'action
d'une distribution sur une fonction test $\varphi$ d'une mani\`ere
purement formelle comme
un produit scalaire entre $\varphi \in \st \subset \lt$
et une fonction qui n'appartient pas \`a $\lt$ :
\begin{eqnarray*}
l_p (\varphi ) \! & \! = \! & \! \left\langle \psi_p ,\varphi \right\rangle
_{L^2} \equiv \int_{\br} dx \, \overline{\psi_p (x)} \, \varphi (x)
\\
\delta_{x_0} (\varphi )
\! & \! = \! & \!
\left\langle \delta_{x_0} , \varphi \right\rangle _{L^2}
\equiv \int_{\br} dx \, \overline{\delta_{x_0} (x)} \, \varphi (x)
\ \ .
\end{eqnarray*}
 
D'une mani\`ere g\'en\'erale, consid\'erons un op\'erateur
auto-adjoint $A$ sur l'espace de Hilbert $\hs$. Les
fonctions propres associ\'ees aux \'el\'ements
du spectre continu de $A$ n'appartiennent pas \`a l'espace
de Hilbert $\hs$ : il faut munir $\hs$ avec un sous-espace
dense appropri\'e $\Omega$ et son dual
$\Omega^{\prime}$ qui contient les vecteurs
propres g\'en\'eralis\'es de $A$, 
\[
\Omega \subset \hs \subset \Omega^{\prime}
\ \ .
\]
Le choix du sous-espace $\Omega$ est \'etroitement li\'e au domaine
de d\'efinition de l'op\'erateur $A$ que l'on souhaite
\'etudier. Alors que l'introduction de l'espace $\Omega$ est n\'ecessaire
pour avoir un probl\`eme math\'ematique bien pos\'e,
celle de $\Omega^{\prime}$ est tr\`es commode, mais elle n'est pas
indispensable pour la d\'etermination du spectre de $A$.
En effet, il existe plusieurs {\em caract\'erisations
du spectre} qui ne font pas appel \`a une extension
de l'espace de Hilbert\footnote{Citons \`a ce sujet les auteurs de
la r\'ef\'erence \cite{rs} : ``We only recommend the abstract
rigged space approach to readers with a strong emotional attachment
to the Dirac formalism." Cette affirmation un peu provocatrice
refl\`ete assez bien l'approche suivie dans  la plupart des ouvrages
d'analyse fonctionnelle.}. Citons trois exemples.
Les diff\'erentes parties du spectre de $A$
peuvent \^etre d\'ecrites par diff\'erentes propri\'et\'es
de la {\em r\'esolvante} $R_A (z) = (A - z {\bf 1})^{-1}$
(o\`u $z \in \bc$) \cite{sg,bgc} ou bien (dans le cas o\`u $A$ est
auto-adjoint) par les propri\'et\'es des
{\em projecteurs spectraux} $E_A (\lambda)$ (o\`u $\lambda \in \br$)
associ\'es \`a $A$
\cite{jvn, sg, rs} ou encore en rempla\c cant
la notion de fonction propre distributive de $A$ par celle de
{\em fonction propre approxim\'ee}
\cite{bgc}. (Cette derni\`ere approche
refl\`ete le fait bien connu
que les distributions telles $\delta_{x_0}$ peuvent \^etre
approxim\'ees arbitrairement bien par des fonctions ordinaires
et continues.)
 
Comme indiqu\'e en haut,
les exemples discut\'es dans cette
section illustrent \`a la fois les probl\`emes pos\'es par un
spectre non born\'e et ceux pos\'es 
par une partie continue
du spectre.
Nous soulignons que ces probl\`emes
ne sont pas reli\'es entre eux et
consid\'erons \`a ce sujet l'exemple
d'une particule
\`a une dimension avec conditions aux limites p\'eriodiques,
c'est-\`a-dire des fonctions d'onde appartenant \`a
l'espace de Hilbert
$L^2 ( [a,b],dx)$ avec
$-\infty < a < b < +\infty$ et satisfaisant aux conditions aux limites
$\psi (a) = \psi (b)$. Dans ce cas,
l'op\'erateur de position (\ref{pos}) admet un spectre continu et
born\'e, donn\'e par l'intervalle
$[a,b]$, alors que le spectre de l'op\'erateur d'impulsion
$\ds{\hbar \over \ri} \ds{d \over dx}$ est discret et non born\'e
(ce qui veut dire que l'impulsion peut seulement prendre certaines
valeurs discr\`etes, mais arbitrairement grandes).
 
 \newpage 
 
\section{Surprises math\'ematiques en m\'ecanique quantique}
 
Des exemples math\'ematiquement simples seront suivis
d'exemples plus sophistiqu\'es et plus int\'eressants du
point de vue de la physique.
Tous ces exemples seront
formul\'es dans le cadre
de la m\'ecanique ondulatoire. Cette th\'eorie \'etant \'equivalente
aux autres formulations de la m\'ecanique quantique, les probl\`emes
mentionn\'es sont aussi pr\'esents dans les autres formulations,
mais \'eventuellement ils y sont moins apparents. Nous uti\-lisons
le langage math\'ematique standard des ouvrages
de m\'ecanique quantique.
La solution des probl\`emes soulev\'es est implicite
dans l'annexe pr\'ec\'edente, mais pour \^etre complet,
nous la d\'etaillerons dans l'annexe suivante tout en faisant appel
aux notions math\'ematiques appropri\'ees.
 
\bigskip
 
%\noindent
{\bf (1)}
Pour une particule \`a une dimension, les op\'erateurs
d'impulsion et de position
$P$ et $Q$ satisfont \`a la relation de commutation canonique
de Heisenberg,
\begin{equation}
\label{hei}
[ P, Q ] = \ds{\hbar \over \ri} \, {\bf 1}
\ \ .
\end{equation}
En prenant la trace de cette relation, on trouve un r\'esultat nul
pour le membre de gauche,
${\rm Tr} \, [P,Q]=0$, alors que
${\rm Tr} \,
( \ds{\hbar \over \ri} \, {\bf 1} ) \neq 0$. Conclusion?
 
\bigskip
 
%\noindent
{\bf (2)}
Consid\'erons des fonctions d'onde $\varphi$ et $\psi$
de carr\'e sommable sur $\br$ ainsi que
l'op\'erateur d'impulsion $P=
\ds{\hbar \over \ri} \, \ds{d \over dx}$.
Une int\'egration par parties donne
\[
\int_{-\infty}^{+\infty} dx \,  \overline{\varphi (x)} \,
(P \psi ) (x ) \ = \
\int_{-\infty}^{+\infty} dx \,
\overline{(P\varphi )(x)} \, \psi (x)   \, + \,
\ds{\hbar \over \ri} \, \left[ \left(
\overline{\varphi} \, \psi \right) (x) \right] _{-\infty}^{+\infty}
\ \ .
\]
Comme $\varphi$ et $\psi$ sont de carr\'e sommable, on conclut
d'habitude que ces fonctions
s'annulent pour $x \to \pm \infty$. Ainsi le dernier terme dans
l'\'equation pr\'ec\'edente est z\'ero et
l'op\'erateur
$P$ est donc hermitien. 

Cependant les ouvrages de math\'ematiques
nous apprennent que les fonctions de carr\'e sommable 
n'admettent en g\'en\'eral pas de  limite pour $x \to \pm \infty$ et
qu'elles ne s'annulent donc pas n\'ecessairement \`a l'infini. 
(Pour illustrer la probl\'ematique,  
nous donnons un exemple \cite{go}
d'une  fonction qui est continue,
positive et sommable sur $\br$ sans pour autant s'annuler
pour $x \to \pm \infty$ : consid\'erons 
$f(x) = \sum_{n=1}^{\infty} f_n(x)$ o\`u $f_n$ est z\'ero sur
$\br$, sauf sur un intervalle de largeur $\ds{2 \over n^2}$
centr\'e en $n$, o\`u le graphe de $f_n$ est un triangle
sym\'etrique par rapport \`a $n$ et de hauteur $1$. L'aire
de ce triangle \'etant \'egale \`a $\ds{1 \over n^2}$,
on a
\[
\int_{-\infty}^{+\infty} dx \, f(x)  =
\sum_{n=1}^{\infty} \, \ds{1 \over n^2} \, < \infty
\ \ ,
\]
mais la fonction
$f$ ne s'annule pas pour $x \to +\infty$.)
On peut m\^eme trouver des fonctions de carr\'e
sommable sur $\br$ qui ne sont pas born\'ees \`a l'infini
\cite{ri} : un exemple d'une telle fonction est donn\'e par 
$f(x) = x^2 \, {\rm exp}\, (-x^8\, {\rm sin}^2 \, x)$, 
ce qui correspond essentiellement \`a un raffinement
de l'exemple pr\'ec\'edent.
Est-ce que l'op\'erateur $P$ est quand m\^eme hermitien
et pourquoi?
\bigskip
 
%\noindent
{\bf (3)}
Consid\'erons les op\'erateurs
$P= \ds{\hbar \over \ri} \, \ds{d \over dx}$ et `$Q=$ multiplication par
$x$'
agissant sur les fonctions d'onde d\'ependant de $x \in \br$.
Comme $P$ et $Q$ sont des op\'erateurs hermitiens, l'op\'erateur
$A =  PQ^3 +Q^3 P$
l'est aussi, car son adjoint est donn\'e par
\[
A^{\dag} = ( PQ^3 +Q^3 P )^{\dag} =
Q^3 P + PQ^3 = A
\ \ .
\]
En cons\'equence, toutes les valeurs propres de $A$ sont r\'eelles.
Pourtant,
on v\'erifie ais\'ement que
\begin{equation}
\label{fon}
Af = \ds{\hbar \over \ri} \, f
\qquad {\rm avec} \quad
f(x) \ =
\left\{
\begin{array}{ll}
\ds{1\over \sqrt{2}} \, |x|^{-3/2} \; {\rm exp} \,
\left( {-1 \over 4x^2} \right)
& {\rm pour} \ x \neq 0
\\
0
& {\rm pour} \ x = 0 \ \ ,
\end{array}
\right.
\end{equation}
ce qui veut dire que $A$ admet la valeur propre complexe
$\hbar / \ri$.
Notons que la fonction $f$ est d\'erivable une infinit\'e de fois
sur $\br$ et qu'elle est de carr\'e sommable, car
\[
\int_{-\infty}^{\infty} dx \, |f(x)|^2 \, = \, 2
\int_0^{\infty} dx \, |f(x)|^2 \, = \,
\int_0^{\infty} dx \, x^{-3} e^{-1/(2x^2)} \, = \,
\left[
e^{-1/(2x^2)} \right]_0^{\infty} \, = \, 1
\ \ .
\]
O\`u est l'erreur?
 
\bigskip
 
%\noindent
{\bf (4)}
Nous consid\'erons une particule enferm\'ee dans l'intervalle
$[0,1]$ et d\'ecrite par une fonction d'onde $\psi$
satisfaisant aux conditions aux limites $\psi(0) =0=\psi(1)$.
L'op\'erateur d'impulsion $P=\ds{\hbar \over \ri} \,
\ds{d \over dx}$ est alors hermitien, car le terme
de surface intervenant dans l'int\'egration par parties
s'annule :
\begin{equation}
\label{itg}
\int_0^1 dx \, \left( \overline{\varphi} \,
(P\psi) -
(\overline{P\varphi} ) \, \psi \right) (x)
=
\ds{\hbar \over \ri} \, \left[ \left(
\overline{\varphi} \, \psi \right) (x) \right] _{0}^{1} = 0
\ \ .
\end{equation}
Comme $P$ est hermitien, ses valeurs propres sont r\'eelles.
Pour d\'eterminer celles-ci,
nous notons que l'\'equation aux valeurs propres,
\[
(P \psi_p)(x) = p \,  \psi_p(x)
\qquad (p \in {\bf R} \ , \ \psi_p \not \equiv 0)
\ \ ,
\]
est r\'esolue par
$\psi_p (x)= c_p \, {\rm exp} \, ( \ds{\ri \over \hbar} px )$
avec $c_p \in \bc - \{ 0 \}$.
La condition aux limites $\psi_p (0) =0$ implique alors
$\psi_p \equiv 0$ et $P$ n'admet donc pas de valeurs propres.
N\'eanmoins le spectre
de $P$ est le plan complexe entier
et $P$ ne repr\'esente pas une observable \cite{sg}.
Comment peut-on comprendre ce r\'esultat
qui para\^it \'etonnant?
 
\bigskip
 
%\noindent
{\bf (5)}
Si l'on introduit les coordonn\'ees polaires dans le plan
ou les coordonn\'ees sph\'eriques dans l'espace, alors
l'angle polaire $\varphi$ et la composante $L_z$
du moment angulaire sont des variables canoniquement
conjugu\'ees en m\'ecanique classique.
En th\'eorie quantique, la variable $\varphi$ devient l'op\'erateur
de multiplication de la fonction d'onde $\psi (\varphi)$ par $\varphi$ et
$L_z = \ds{\hbar \over \ri} \, \ds{\pa \over \pa \varphi}$,
ce qui implique la relation de commutation
\begin{equation}
\label{lzp}
[ L_z ,  \varphi ] = {\hbar \over \ri} \, {\bf 1}
\ \ .
\end{equation}
Ces op\'erateurs agissant sur des fonctions d'onde p\'eriodiques
($\psi (0) = \psi (2\pi)$)
sont hermitiens. Par ailleurs,
$L_z $ admet un syst\`eme complet de fonctions propres
orthonorm\'ees $\psi_m$,
\begin{equation}
L_z \psi_m =  m \hbar \, \psi_m
\qquad {\rm avec} \quad
\psi_m (\varphi ) = {1 \over \sqrt{2 \pi}} \;
{\rm exp} \, (\ri
m \varphi )
\quad {\rm et} \quad
m \in {\bf Z}
\ \ .
\end{equation}
(Pour les fonctions d'onde $\psi$, nous
sp\'ecifions seulement la d\'ependance de la variable angulaire
$\varphi$ et pour l'orthonormalisation,
nous nous r\'ef\'erons au produit scalaire standard des
fonctions de carr\'e sommable sur l'intervalle $[0, 2\pi )$ :
\[
\langle \psi_1 , \psi_2 \rangle = \int_0^{2 \pi}
d\varphi \ \overline{\psi_1 (\varphi )} \, \psi_2 (\varphi )
\ \ .)
\]
En prenant la valeur moyenne de l'op\'erateur
$[ L_z ,  \varphi ]$ dans l'\'etat $\psi_m$ \cite{car,grau}
et en tenant compte du fait que $L_z $ est hermitien,
on trouve que
\begin{eqnarray}
{\hbar \over \ri} \ = \
\langle \psi_m ,
{\hbar \over \ri} \, {\bf 1} \, \psi_m \rangle
& \stackrel{(\ref{lzp})}{=} &
\langle \psi_m , L_z  \varphi \, \psi_m \rangle -
\langle \psi_m , \varphi L_z \, \psi_m \rangle
\nonumber \\
& = &
\langle L_z ^{\dag} \, \psi_m , \varphi \, \psi_m \rangle -m \hbar
\, \langle \psi_m , \varphi \, \psi_m \rangle
\label{deriv}
\\
& = &
( m \hbar -  m \hbar )
\, \langle \psi_m , \varphi \, \psi_m \rangle
\ = \ 0
\ \ .
\nonumber
\end{eqnarray}
Il doit y avoir un petit probl\`eme quelque part...
 
\bigskip
 
%\noindent
{\bf (6)}
Rajoutons un peu \`a la confusion de l'exemple pr\'ec\'edent!
En 1927, Pauli a not\'e que
la relation de commutation canonique (\ref{hei})
implique la relation d'incertitude de Heisenberg
$\Delta P \cdot \Delta Q  \geq \ds{\hbar \over 2}$
en vertu
de l'in\'egalit\'e de Cauchy et Schwarz. Comme la relation
de commutation (\ref{lzp}) a la m\^eme forme que (\ref{hei}),
on peut d\'eduire de la m\^eme mani\`ere la relation
d'incertitude
\begin{equation}
\label{inc}
\Delta L_z \cdot \Delta \varphi \geq \ds{\hbar \over 2}
\ \ .
\end{equation}
Le raisonnement physique suivant montre que cette in\'egalit\'e
ne peut pas \^etre correcte \cite{ju, car, gap}.
On peut toujours trouver un \'etat pour lequel
$\Delta L_z  < \hbar /  4\pi$
et alors l'incertitude sur l'angle $\varphi$
devrait \^etre plus grande que $2 \pi$, ce qui n'a pas de
sens physique, puisque
$\varphi$ prend des valeurs dans
l'intervalle $[0,2 \pi )$.
Comment se fait-il que la relation (\ref{lzp}) est correcte,
mais que la conclusion (\ref{inc}) ne l'est pas?
 
D'ailleurs cet exemple montre que la relation d'incertitude
$\Delta A\cdot \Delta B  \geq \ds{1 \over 2} \,
| \, \langle [ A,B] \rangle \, |$ pour deux observables quelconques
$A$ et $B$ (dont on trouve la d\'erivation dans la plupart des livres de
m\'ecanique quantique) n'est pas valable dans cette g\'en\'eralit\'e.

\bigskip
 
%\noindent
{\bf (7)}
Consid\'erons une particule de masse $m$ dans le puits de potentiel
infini
\[
V(x) = \left\{
\begin{array}{ll}
0
& {\rm si} \ \; |x| \leq a \quad  \ (a >0)
\\
\infty
& {\rm sinon} \ .
\end{array}
\right.
\]
L'hamiltonien pour 
la particule enferm\'ee dans le puits 
est simplement $H = \ds{-\hbar^2 \over 2m}
\ds{d^2 \over dx^2}$.
Soit
\begin{equation}
\label{psi}
\psi(x) =
\ds{\sqrt{15} \over 4 a^{5/2}} \, (a^2 - x^2) \qquad
{\rm pour} \ |x| \leq a \quad  (\, {\rm et} \ \, \psi (x) =0 \ \,
{\rm autrement} \, )
\end{equation}
la fonction d'onde norm\'ee de la particule \`a un instant donn\'e.
Comme $H^2 \psi =
\ds{\hbar^4 \over 4 m^2} \, \ds{d^4 \psi \over dx^4} =0$,
la valeur moyenne de l'op\'erateur $H^2$
dans l'\'etat $\psi$
s'annule :
\begin{equation}
\label{caf}
\langle H^2 \rangle_{\psi} = \langle \psi, H^2 \psi \rangle
= \int_{-a}^{+a} dx \; \overline{\psi (x)} \, (H^2  \psi)(x)  = 0
\ \ .
\end{equation}
Cette valeur moyenne peut aussi \^etre d\'etermin\'ee \`a partir
des valeurs et fonctions propres de $H$,
\begin{equation}
\label{pui}
H \varphi_n = E_n \varphi_n
\qquad {\rm avec} \qquad
E_n = \ds{\pi^2 \hbar^2 \over 8ma^2} \, n^2
\qquad (n=1,2,...)
\ \ ,
\end{equation}
en appliquant la formule
\begin{equation}
\langle H^2 \rangle_{\psi} = \sum_{n=1}^{\infty} E_n^2 \, p_n
\qquad {\rm avec} \ \, p_n = | \langle \varphi_n ,\psi \rangle | ^2
\ \ .
\end{equation}
En proc\'edant de cette mani\`ere,
on ne trouve certainement pas un r\'esultat nul,
car $E_n^2 >0$ et $0\leq p_n \leq 1, \, \, \sum_{n=1}^{\infty} p_n =1$.
En fait, le calcul donne $\langle H^2 \rangle_{\psi}
= \ds{15\hbar^4 \over 8m^2 a^4}$. Lequel des deux
r\'esultats est correct et d'o\`u provient l'incoh\'erence? \cite{grau}

 \newpage 
 
\section{Il n'y a pas de surprise}
 
La r\'esolution des probl\`emes et contradictions apparentes
de l'annexe pr\'ec\'edente peut \^etre paraphras\'ee de la mani\`ere
suivante \cite{rs} : la th\'eorie des op\'erateurs lin\'eaires sur des
espaces vectoriels de dimension infinie est plus compliqu\'ee
et plus int\'eressante que la th\'eorie des matrices de dimension finie.
Discutons maintenant  
les probl\`emes mentionn\'es tout en appliquant les r\'esultats 
math\'ematiques de l'annexe A. 
 
\bigskip
 
{\bf (1)}
Supposons que la relation de commutation
$[P,Q] = \ds{\hbar \over \ri} \, {\bf 1}$ soit satisfaite par des
op\'erateurs
$P$ et $Q$ agissant sur un espace de Hilbert $\hs$ de dimension finie
$n$ (c'est-\`a-dire $\hs \simeq {\bf C}^n$). 
Dans ce cas, $P$ et $Q$ peuvent \^etre
r\'ealis\'es par des matrices carr\'ees $n\times n$,
la trace est une
op\'eration bien d\'efinie et nous obtenons le r\'esultat
\[
0 = {\rm Tr} \, [P,Q ] 
\stackrel{(\ref{hei})}{=} 
{\rm Tr} \, ( \ds{\hbar \over \ri}
{\bf 1}_n ) =
\ds{\hbar \over \ri}\; n
\ \ .
\]
On en d\'eduit que la relation de Heisenberg ne peut pas \^etre
r\'ealis\'ee
sur un espace de Hilbert de dimension finie. La
m\'ecanique quantique doit donc \^etre
formul\'ee sur un espace de Hilbert de dimension infinie :
sur un tel espace, 
la trace n'est plus une op\'eration bien d\'efinie pour tous les
op\'erateurs
(en particulier, la trace de l'op\'erateur ${\bf 1}$ n'existe pas) et
on ne peut donc 
plus d\'eduire de contradiction de la relation de commutation de
Heisenberg de la mani\`ere indiqu\'ee.

Une incoh\'erence peut encore \^etre d\'eduite
d'une autre fa\c con sur un espace de Hilbert de dimension infinie 
en supposant que $P$ et $Q$ 
 sont tous les deux des op\'erateurs born\'es \cite{rs}; 
 par cons\'equence,  
 au moins l'un des deux op\'erateurs $P$ et $Q$ satisfaisant 
 la relation de Heisenberg doit \^etre non born\'e 
 et cette relation fondamentale ne peut donc pas \^etre discut\'ee 
 sans se soucier des domaines de d\'efinition des op\'erateurs.

\bigskip
 
{\bf (2)}
Le domaine de d\'efinition maximal de l'op\'erateur
$P= \ds{\hbar \over \ri} \, \ds{d \over dx}$ sur l'espace de Hilbert
$\lt$
est\footnote{Comme l'int\'egrale intervenant dans la d\'efinition de
l'espace
$\lt$ est celle de Lebesgue, il faut seulement s'assurer que
les fonctions consid\'er\'ees
se comportent correctement `presque partout' par
rapport \`a la mesure de Lebesgue (voir livres d'analyse) :
$\psi^{\prime} \in \lt$ veut donc dire que la d\'eriv\'ee
$\psi^{\prime}$ existe presque partout et qu'elle appartient \`a $\lt$.}
\[
{\cal D}_{\rm max} (P) = \{ \psi \in \lt \, | \,
\psi^{\prime} \in \lt \}
\ \ .
\]
Les fonctions appartenant \`a ${\cal D}_{\rm max} (P)$ poss\`edent
donc certaines propri\'et\'es
de r\'egularit\'e et leur d\'eriv\'ee est de carr\'e sommable sur
$\br$. En particulier, ces fonctions sont continues
et leur
limite pour $x \to \pm \infty$ 
est z\'ero \cite{ri, bgc}, ce qui implique que l'op\'erateur $P$
agissant sur ${\cal D}_{\rm max} (P)$
est hermitien.
La fonction non born\'ee \`a l'infini que nous
avons mentionn\'ee
 est d\'erivable,
mais sa d\'eriv\'ee n'est pas de carr\'e sommable et elle
n'appartient donc pas \`a ${\cal D}_{{\rm max}}(P)$.
 
Un autre domaine de d\'efinition acceptable pour $P$
est l'espace de Schwartz ${\cal S}(\br)
\subset {\cal D}_{{\rm max}} (P)$.
Dans ce cas, les
fonctions sur lesquelles agit l'op\'erateur $P$
ont m\^eme une d\'ecroissance rapide \`a
l'infini.
 
\bigskip
 
{\bf (3a)}
L'espace de Schwartz $\st \subset \lt$ est un domaine de d\'efinition
invariant pour les op\'erateurs $P$ et $Q$ et donc aussi pour
$A= PQ^3 +Q^3 P$ :
\[
A : \st \longrightarrow \st
\ \ .
\]
Une int\'egration par parties montre que l'op\'erateur $A$ 
ainsi d\'efini est {\em
hermitien} : \[
\langle g , A f \rangle  =
\langle A g , f \rangle
\qquad \mbox{pour tout} \ f,g \in {\cal D}(A) = \st
\ \ .
\]
La fonction $f$ donn\'ee par (\ref{fon}) appartient \`a
l'espace de Hilbert $\lt$,
mais elle n'appartient
pas au domaine de d\'efinition de $A$, puisqu'elle
ne d\'ecro\^it pas plus rapidement que l'inverse d'un polyn\^ome
quelconque
\`a l'infini : par exem\-ple,
$x^3 f(x) \propto  x^{3/2} \, {\rm exp}\, [-1 /(4x^2)]$ n'est pas
born\'e pour $x \to +\infty$.
En cons\'equence, $\hbar /\ri$ n'est {\em pas} une valeur propre de $A$.
 
Par contre, $\hbar /\ri$ est une valeur propre de $A^{\dag}$
\cite{blt}.
Avant de discuter ce point, il est pr\'ef\'erable de
r\'esoudre d'abord les autres probl\`emes.
 
\bigskip
 
{\bf (4a)}
Les r\'esultats \'etonnants que nous avons cit\'es dans cet exemple
indiquent
qu'il ne suffit pas de v\'erifier qu'un op\'erateur est hermitien
pour l'identifier comme une obser\-vable : ceci est bien connu
\cite{d, ct}. Par ailleurs, ces r\'esultats indiquent
que le spectre d'un op\'erateur n'est pas
simplement l'ensemble de ses valeurs propres (comme
c'est le cas pour les matrices
de dimension finie). Dans la suite, nous explicitons ces deux points.
 
Le domaine de d\'efinition que l'on consid\`ere ici pour
l'op\'erateur $P$ sur $\hs =  L^2
([0,1], dx)$ est
\begin{equation}
\label{dir}
{\cal D} (P) = \{ \psi \in \hs \, | \,
\psi^{\prime} \in \hs \ {\rm et} \ \psi(0) = 0 = \psi(1) \}
\ \ .
\end{equation}
Sur ce domaine, $P$ est hermitien :
\[
\langle \varphi , P\psi \rangle =
\langle P\varphi , \psi \rangle
\qquad \mbox{pour tout} \ \varphi, \psi \in {\cal D} (P)
\ \ .
\]
Comme il n'existe pas de solution de l'\'equation aux valeurs
propres
\[
P \psi_p = p \, \psi_p \qquad
{\rm avec} \ \, \psi_p \in {\cal D}(P) \ \, {\rm et} \ \,
\psi_p \not \equiv 0
\ \ ,
\]
l'op\'erateur $P$
n'admet aucun vecteur propre 
(et pas non plus de vecteur propre g\'en\'eralis\'e).
En cons\'equence, il n'existe pas de
syst\`eme complet de vecteurs propres de $P$ et l'op\'erateur
$P$ n'est donc pas une observable selon la d\'efinition
habituelle donn\'ee en m\'ecanique quantique \cite{d, ct}.
En effet, l'op\'erateur $P$ avec le domaine de d\'efinition
(\ref{dir}) est hermitien, mais pas auto-adjoint.
Pour v\'erifier ceci, nous
rappelons de l'annexe A.2 que
le domaine de d\'efinition de $P^{\dag}$
est donn\'ee par
\[
{\cal D} (P^{\dag}) = \{ \varphi \in \hs \, | \, \exists\,
\tilde{\varphi} \in \hs \ \mbox{tel que} \
\langle \varphi , P \psi \rangle =
\langle \tilde \varphi , \psi \rangle
\ \mbox{pour tout} \ \psi  \in {\cal D} (P) \}
\]
et que la prescription d'op\'eration de $P^{\dag}$ est
d\'etermin\'ee par la relation
\begin{equation}
\langle \varphi , P \psi \rangle =
\langle P^{\dag} \varphi , \psi \rangle
\qquad \mbox{pour tout} \ \psi \in {\cal D}(P)
\ \ .
\end{equation}
L'int\'egration par parties (\ref{itg})
ou, plus pr\'ecisement,
\[
\int_0^1 \! d x \, (  \overline{\varphi} \, P\psi
-
\left( \overline{\ds{\hbar \over \ri} \,
\ds{d\varphi \over dx} }  \right) \psi ) ( x )
= \ds{\hbar \over \ri} \left[ \overline{\varphi( 1)} \psi (1) -
\overline{ \varphi ( 0)} \psi (0) \right] =0
\ \ \ \mbox{pour tout} \ \, \psi \in {\cal D}( P)
\]
montre
que les conditions aux limites satisfaites par
$\psi \in {\cal D}(P)$ suffisent d\'ej\`a pour annuler
le terme de surface
et que $P^{\dag}$ agit de la m\^eme mani\`ere
que $P$. Ainsi
\[
P^{\dag}  = \ds{\hbar \over \ri} \, \ds{d \over dx}
\quad , \quad
{\cal D} (P^{\dag}) = \{ \varphi \in \hs \, | \, \varphi^{\prime}
\in \hs \}
\ \ .
\]
Le domaine de d\'efinition de $P^{\dag}$ est donc plus large
que celui de $P$ :
${\cal D}(P) \subset {\cal D}(P^{\dag})$.
Nous en concluons que
$P$ est hermitien , mais pas auto-adjoint :
$P \neq P^{\dag}$, car
${\cal D}(P) \neq {\cal D}(P^{\dag})$.
 Le spectre de $P$ sera discut\'e plus loin.  
 
\bigskip
 
{\bf (5)}
L'op\'erateur de multiplication par $\varphi$
sur l'espace de Hilbert
$\hs = L^2
([0,2 \pi ], d\varphi )$ est partout d\'efini et auto-adjoint :
\[
\langle g , \varphi \, f \rangle
 = \langle \varphi \, g , f \rangle \qquad \mbox{pour tout} \ \
g,f \in \hs
\ \ .
\]
La discussion
pr\'ec\'edente concernant l'op\'erateur $P = \ds{\hbar \over \ri} \,
\ds{d \over dx}$ sur
$L^2
([0,1], dx )$ s'applique verbatim \`a l'op\'erateur
$L_z = \ds{\hbar \over \ri} \,
\ds{d \over d\varphi}$ sur
$L^2
([0,2 \pi ], d\varphi )$ : une int\'egration par parties donne
\begin{equation}
\label{uti}
\int_0^{2\pi} \! d \varphi \, (  \overline{g} \, L_z f
-
\left( \overline{\ds{\hbar \over \ri} \,
\ds{dg \over d\varphi} }  \right)  f ) (\varphi )
= \ds{\hbar \over \ri} \left[ \overline{g(2 \pi)} f (2 \pi) -
\overline{g( 0)} f (0) \right]
\quad \mbox{pour tout} \ \, f \in {\cal D}(L_z).
\end{equation}
A cause du caract\`ere p\'eriodique de
l'angle polaire, les fonctions appartenant au domaine de d\'efinition
de $L_z$ sont p\'eriodiques\footnote{A ce sujet,
nous remarquons que
l'utilisation des coordonn\'ees polaires
attribue un r\^ole distingu\'e au demi-axe polaire $\varphi =0$,
alors que cet axe n'est pas privil\'egi\'e si l'on choisit
d'autres syst\`emes de coordonn\'ees
comme les coordonn\'ees cart\'esiennes : aussi une discontinuit\'e
des fonctions d'onde sur cet axe ($f(2 \pi)= {\rm e}^{\ri \alpha}
f(0)$ avec
$\alpha \neq 0$)
n'a pas de raison d'\^etre.} :
\begin{equation}
\label{oaa}
L_z = \ds{\hbar \over \ri} \, \ds{d \over d\varphi} \quad , \quad
{\cal D} (L_z ) = \{ f \in \hs \; | \;
f^{\prime} \in \hs \ \; {\rm et} \ \; f(0) = f(2 \pi ) \}
\ \ .
\end{equation}
Par cons\'equent, le terme de surface dans (\ref{uti})
s'annule si et seulement si $g(0) = g(2 \pi)$ :
ceci implique que $L_z^{\dag}$ agit de la m\^eme mani\`ere que
$L_z$ et admet le m\^eme domaine de d\'efinition, donc
l'op\'erateur (\ref{oaa}) est auto-adjoint.
 
Pour d\'eterminer le domaine de d\'efinition du commutateur
$[L_z, \varphi ]$, nous notons que pour deux op\'erateurs $A$ et $B$,
on a
\begin{eqnarray}
{\cal D} (A+B) & = & {\cal D}(A) \cap {\cal D}(B)
\\
{\cal D} (AB) & = & \{ f \in {\cal D}(B)\, | \, Bf \in {\cal D}(A) \}
\nonumber
\ \ .
\end{eqnarray}
Ainsi
${\cal D} ([L_z, \varphi]) = {\cal D}(L_z \varphi ) \cap
{\cal D} (\varphi L_z )$ avec
\begin{eqnarray*}
{\cal D} (\varphi L_z )
& = & \{ f \in {\cal D}(L_z) \, | \, L_z f \in {\cal D}(\varphi )= \hs \}
\ = \ {\cal D} (L_z )
\\
{\cal D} (L_z  \varphi) & = &
\{ f \in {\cal D}(\varphi) =\hs \, | \, \varphi  f \in {\cal D}(L_z) \}
\ \ .
\end{eqnarray*}
Or la fonction $\tilde f \equiv \varphi f$ qui intervient dans la
derni\`ere expression 
prend les valeurs
\begin{eqnarray*}
\tilde f (0) & = & (\varphi f)(0) \ = \ 0
\\
\tilde f (2 \pi ) & = & (\varphi f)(2\pi ) \ = \ 2 \pi \, f (2 \pi)
\end{eqnarray*}
et $\tilde f \in {\cal D}(L_z)$ implique
$\tilde f (0) =
\tilde f (2 \pi )$, donc $f(2\pi ) = 0$.
 
En r\'esum\'e,
\begin{eqnarray}
{\cal D} (\varphi L_z )
& = & {\cal D} (L_z )
\nonumber
\\
{\cal D} (L_z  \varphi) & = &
\{ f \in \hs \, | \, f^{\prime} \in \hs \ \ {\rm et} \ \
f (2 \pi ) =0 \}
\\
{\cal D} ([ L_z , \varphi ]) & = &
\{ f \in \hs \, | \, f^{\prime} \in \hs \ \ {\rm et} \ \
f (0) =0 =
f (2 \pi ) \}
\ \ .
\nonumber
\end{eqnarray}
Les fonctions propres
$\psi_m ( \varphi ) = \ds{1 \over \sqrt{2\pi} } \, {\rm exp}\,
( \ri  m \varphi )$ de $L_z$
n'appartiennent pas au domaine de
d\'efinition de
$[ L_z , \varphi ]$, puisqu'elles ne s'annulent pas en $0$ et
$2\pi$ : ainsi
la d\'erivation (\ref{deriv}) n'a pas
de sens.
 
\bigskip
 
{\bf (6)}
Consid\'erons deux observables $A,B$ (op\'erateurs auto-adjoints
sur un espace de Hilbert $\hs$)
et un \'etat $\psi$ (vecteur de norme $1$ appartenant \`a $\hs$).
La relation d'incertitude pour $A,B$ est d'habitude \'ecrite
sous la forme \cite{gap}
\begin{equation}
\label{ri}
\Delta_{\psi} A \cdot
\Delta_{\psi} B \geq \ds{1 \over 2} \, \mid \langle \psi ,
\ri [A, B] \psi \rangle \mid
\ \ ,
\end{equation}
o\`u
$(\Delta_{\psi} A)^2 = \| (A -
\langle A\rangle_{\psi} {\bf 1} ) \psi \|^2$
avec
$\langle A\rangle_{\psi} = \langle \psi , A\psi \rangle$ et de m\^eme
pour $B$.
Ainsi le membre de gauche de la relation (\ref{ri}) est d\'efini
pour $\psi \in {\cal D}(A) \cap {\cal D}(B)$
(qui est
pr\'ecisement le sous-espace de $\hs$
contenant tous les \'etats $\psi$
pour lesquels les incertitudes
$\Delta_{\psi} A$ et
$\Delta_{\psi} B$ ont toutes les deux une signification physique).
Par contre le membre de droite est seulement d\'efini sur le
sous-espace
${\cal D} ([A,B]) =
{\cal D} (AB) \cap
{\cal D} (BA)$ qui est en g\'en\'eral beaucoup plus petit.
 
Mais $A,B$ \'etant auto-adjoints, la relation (\ref{ri}) peut
\^etre r\'e\'ecrite sous la forme \cite{krau}
\begin{equation}
\label{rir}
\Delta_{\psi} A \cdot
\Delta_{\psi} B \geq \ds{1 \over 2} \, \mid \ri \langle A\psi ,
B \psi \rangle - \ri \langle B\psi , A \psi \rangle \mid
\ \ ,
\end{equation}
o\`u le domaine de d\'efinition
du membre de droite est maintenant le m\^eme
que celui de gauche, c'est-\`a-dire
${\cal D}(A) \cap {\cal D}(B)$.
Ainsi le produit des incertitudes
pour deux observables $A$ et $B$ n'est pas d\'etermin\'ee par leur
commutateur, mais par la forme hermitienne 
sesquilin\'eaire\footnote{i.e. 
 $\Phi_{A,B} (f,g)$ est lin\'eaire en $g$, 
 antilin\'eaire en $f$
  et $\Phi_{A,B} (g,f) = \overline{\Phi_{A,B} (f,g)}$.}
\[
\Phi_{A,B} (f,g) =
\ri \langle A f, Bg \rangle
-\ri \langle B f , A g \rangle
\qquad \mbox{pour tout} \ \,  f,g \in
{\cal D}(A) \cap {\cal D}(B)
\ \ .
\]
La d\'erivation de l'in\'egalit\'e
(\ref{rir}) est la m\^eme que celle de (\ref{ri}) (voir par exemple
\cite{af} pour cette derni\`ere) et se fait en quelques
lignes : soit $\psi \in {\cal D}(A) \cap {\cal D} (B)$ et soit
\[
\hat A = A - \langle A \rangle _{\psi} {\bf 1}
\qquad , \qquad
\hat B = B - \langle B \rangle _{\psi} {\bf 1}
\ \ ;
\]
en utilisant le fait que $A$ et $B$ sont auto-adjoints et en appliquant
l'in\'egalit\'e triangulaire et celle de Cauchy et Schwarz,
nous trouvons l'in\'egalit\'e (\ref{rir}) :
\begin{eqnarray*}
\mid \ri \langle A \psi , B \psi  \rangle
-\ri \langle B \psi , A \psi \rangle \mid
& = &
\mid \ri \langle \hat A \psi , \hat B \psi  \rangle
-\ri \langle \hat B \psi , \hat A \psi \rangle \mid
\\
& \leq &
\mid  \langle \hat A \psi , \hat B \psi  \rangle  \mid +
\mid  \langle \hat B \psi , \hat A \psi  \rangle  \mid
\ = \ 2 \,
\mid  \langle \hat A \psi , \hat B \psi  \rangle  \mid
\\
& \leq &  2 \,
\| \hat A \psi \| \cdot
\| \hat B \psi \|  \ = \
2\, \Delta_{\psi} A \cdot
\Delta_{\psi} B
\ \ .
\end{eqnarray*}

Montrons
maintenant que cette modification n'est pas purement cosm\'etique.
Pour $A=P= \ds{\hbar \over \ri} \, \ds{d \over dx}$ et $B=Q=x$
sur $\hs = L^2 ( {\bf R}, dx)$, le membre de droite
de l'in\'egalit\'e (\ref{rir})
est facile \`a \'evaluer par int\'egration par parties
et implique la relation d'incertitude bien connu
$\Delta_{\psi} P \cdot
\Delta_{\psi} Q \geq \ds{\hbar  \over 2}$ pour $\psi \in
{\cal D}(P) \cap {\cal D}(Q)$.
Par contre, pour $A= L_z = \ds{\hbar \over \ri} \, \ds{d \over
d\varphi}$ et $B=\varphi$
sur $\hs = L^2 ( [0,2\pi ] , d\varphi )$, le terme de surface
intervenant dans l'int\'egration par parties ne s'annule pas et
conduit \`a la relation d'incertitude
\begin{equation}
\label{lphi}
\Delta_{\psi} L_z \cdot
\Delta_{\psi} \varphi \geq \ds{\hbar  \over 2}
\ \mid 1 - 2 \pi | \psi (2 \pi ) | ^2 \mid \qquad
\mbox{pour tout} \ \; \psi \in {\cal D}(L_z) \cap {\cal D}(\varphi)
= {\cal D} (L_z)
.
\end{equation}
Ainsi le produit des incertitudes
$\Delta_{\psi} L_z$ et
$\Delta_{\psi} \varphi$ peut devenir plus petit que $\hbar /2$
- voir Galindo et Pascual \cite{gap} pour un exemple.
 Si $\psi \in {\cal D} ([L_z , \varphi ] )$, c'est-\`a-dire
$\psi (2 \pi )=0$, l'in\'egalit\'e (\ref{ri}) peut aussi \^etre 
appliqu\'ee 
et elle donne le m\^eme r\'esultat que (\ref{lphi}).
 
Alors que l'in\'egalit\'e (\ref{lphi}) est math\'ematiquement correcte,
elle n'est pas acceptable sous cette forme du point de vue physique :
si l'on d\'efinit la valeur moyenne et l'incertitude de l'observable
$\varphi$ par les formules habituelles, ces expressions n'ont pas
les bonnes propri\'et\'es de transformation par rapport aux rotations
$\psi (\varphi ) \to ({\rm exp} \, ({\ri \over \hbar}
 \alpha L_z ) ) \psi) (\varphi )
= \psi (\varphi + \alpha)$. Nous renvoyons \`a la litt\'erature
\cite{ju,krau}
pour une l\'eg\`ere modification de (\ref{lphi})
qui tient compte de ce probl\`eme ainsi que pour des estimations
du produit
$\Delta_{\psi} L_z \cdot
\Delta_{\psi} \varphi$ qui ne d\'ependent pas explicitement
de l'\'etat particulier $\psi$ que l'on consid\`ere.
Des probl\`emes similaires concernant les op\'erateurs de
phase et du nombre, qui sont d'int\'er\^et en optique
quantique, sont discut\'es dans \cite{car}. 
 
\bigskip
 
{\bf (7)}
Une solution purement formelle du probl\`eme peut \^etre
obtenue, si l'on consid\`ere la fonction d'onde comme d\'efinie 
sur tout l'axe
r\'eel au lieu de se limiter \`a l'intervalle $[-a, +a]$.
En effet, pour la fonction $\psi$ d\'efinie par (\ref{psi}),
la discontinuit\'e de
$\psi^{\prime \prime}$ en $x = \pm a$ implique que
$\psi^{\prime \prime \prime \prime}$ est donn\'e par des
d\'eriv\'ees de la fonction g\'en\'eralis\'ee de Dirac :
\[
\psi^{\prime \prime \prime \prime} (x) = -
{\sqrt{15} \over 2 a^{5/2}} \left[
\delta^{\prime} (x+a) - \delta^{\prime} (x-a) \right]
\qquad {\rm pour} \ \, x \in \br
\ \ .
\]
Substitution de cette expression dans $\langle \psi , H^2 \psi \rangle$
conduit alors au m\^eme r\'esultat non nul que l'\'evaluation de
$\sum_{n=1}^{\infty} E_n^2 p_n$.
L'incoh\'erence mentionn\'ee provient donc du fait qu'on n'a
pas correctement tenu compte des conditions aux limites dans
le calcul (\ref{caf}).
 
Dans la suite, nous montrons comment un raisonnement rigoureux
limit\'e \`a l'intervalle $[-a , +a]$ permet d'incorporer
les conditions aux limites et de confirmer le
r\'esultat non nul pour $\langle H^2 \rangle_{\psi}$.
Pour commencer, nous d\'efinissons $H$ et $H^2$ en tant qu'op\'erateurs
auto-adjoints sur l'espace de Hilbert $\hs = L^2 ([-a,+a], dx)$.
 
Le puits de potentiel infini est une id\'ealisation math\'ematique
qui est
\`a interpr\'eter comme la limite $V_0 \rightarrow \infty$
d'un puits de potentiel fini de hauteur $V_0$. Pour ce dernier,
on trouve qu'en dehors du puits, les fonctions d'onde
des \'etats stationnaires tendent vers z\'ero pour
$V_0 \rightarrow \infty$ et par
cons\'equent $\psi(\pm a)=0$ est la condition aux limites appropri\'ee
pour la particule enferm\'ee dans le puits infini. Analysons
maintenant si $H$ est
auto-adjoint si l'on le fait agir sur des fonctions suffisamment
d\'erivables satisfaisant $\psi (\pm a)=0$ : par
deux int\'egrations par parties successives, nous trouvons
\begin{eqnarray*}
\langle \varphi , H \psi \rangle & \equiv &
\ds{- \hbar^2 \over 2m} \int_{-a}^{+a} dx \, \overline{\varphi(x)}
\psi^{\prime \prime}(x)
\\
&= &
\ds{- \hbar^2 \over 2m} \int_{-a}^{+a} dx \, \overline{\varphi^{\prime
\prime}(x)}
\psi (x) +
\ds{\hbar^2 \over 2m}
\left[ (\overline{\varphi^{\prime}} \psi - \overline{\varphi}
\psi^{\prime} )(x) \right]_{-a}^{+a}
\\
& = & \langle H^{\dag} \varphi , \psi \rangle -
\ds{\hbar^2 \over 2m}
\left[ \overline{\varphi} (a)\psi^{\prime} (a) - \overline{\varphi}(-a)
\psi^{\prime} (-a) \right]
\ \ .
\end{eqnarray*}
Comme nous n'avons pas de contraintes sur $\psi^{\prime}(\pm a)$,
le terme de surface s'annule si et seulement si $\varphi (\pm a)=0$.
En r\'esum\'e, $H^{\dag}$ op\`ere de la m\^eme mani\`ere que $H$
et les fonctions $\varphi$ appartenant \`a son
domaine de d\'efinition
satisfont les m\^emes conditions que celles du domaine de $H$.
Donc l'op\'erateur
$H= \ds{-\hbar^2 \over 2m} \, \ds{d^2 \over dx^2}$ agissant sur
$\hs$ avec le domaine de d\'efinition
\begin{equation}
{\cal D} (H) = \{ \psi \in \hs \; \mid \; \psi^{\prime \prime}
\in \hs \ \; {\rm et} \ \; \psi(\pm a) =0 \}
\end{equation}
est un op\'erateur auto-adjoint (une observable).
Son spectre, qui a \'et\'e explicit\'e
dans l'\'equation (\ref{pui}), est discret et non d\'eg\'en\'er\'e
et les fonctions propres associ\'ees sont donn\'ees par
\[
\varphi_n (x) = \left\{
\begin{array}{ll}
\ds{1 \over \sqrt a} \  {\rm sin}  \left( \ds{n\pi \over 2a} x \right)
& \quad {\rm pour} \ \, n=2,4,6,...
\\
\ds{1 \over \sqrt a} \  {\rm cos}  \left( \ds{n\pi \over 2a} x \right)
& \quad {\rm pour} \ \, n=1,3,5,... \ \ .
\end{array}
\right.
\]
Par cons\'equent, la d\'ecomposition spectrale de $H$ s'\'ecrit
$H= \sum_{n=1}^{\infty} E_n {\cal P}_n$ o\`u ${\cal P}_n$
est le projecteur sur l'\'etat normalis\'e $\varphi_n$ :
${\cal P}_n \psi = \langle \varphi_n , \psi \rangle \, \varphi_n$.
 
D'apr\`es le th\'eor\`eme spectral \cite{rs},
l'op\'erateur $H^2$ est d\'efini \`a partir de la d\'ecomposition
spectrale de $H$,
\begin{equation}
\label{h2}
H^2 = \sum_{n=1}^{\infty} E_n^2 \, {\cal P}_n
\ \ ,
\end{equation}
ce qui implique que $H^2 \varphi_n = E_n^2 \varphi_n$.
Pour d\'eterminer explicitement le domaine de d\'efinition sur lequel
cet op\'erateur est auto-adjoint, on fait quatre int\'egrations par
parties successives :
\begin{eqnarray*}
\langle \varphi , H^2 \psi \rangle & \equiv &
\ds{\hbar^4 \over 4m^2} \int_{-a}^{+a} dx \, \overline{\varphi(x)}
\psi^{\prime \prime \prime \prime}(x)
\\
&= &
\ds{\hbar^4 \over 4m^2} \int_{-a}^{+a} dx \, \overline{\varphi^{\prime
\prime \prime \prime}(x)}
\psi (x) +
\ds{\hbar^4 \over 4m^2} \left[
( \overline{\varphi} \psi^{\prime \prime \prime} - \overline{\varphi
^{\prime}} \psi^{\prime \prime} +
\overline{\varphi^{\prime \prime}} \psi^{\prime} -
\overline{\varphi^{\prime \prime\prime}} \psi )(x)
\right]_{-a}^{+a}
.
\end{eqnarray*}
Les conditions aux limites $\psi (\pm a) =0 = \varphi (\pm a)$
du puits infini
enl\`event la premi\`ere et la derni\`ere contribution du
terme de surface. Pour annuler
les autres, il y a diff\'erentes possibilit\'es, par exemple
$\psi^{\prime} (\pm a) =0 = \varphi^{\prime} (\pm a)$ ou
$\psi^{\prime \prime} (\pm a) =0 = \varphi^{\prime \prime} (\pm a)$.
Or, par suite de la d\'efinition (\ref{h2}) de $H^2$, les fonctions
propres $\varphi_n$ de $H$ doivent appartenir au domaine de
$H^2$ : comme ces fonctions satisfont $\varphi_n ^{\prime \prime}
(\pm a)=0$, le domaine
de d\'efinition de l'observable $H^2$ est
\begin{equation}
\label{dh2}
{\cal D} (H^2) = \{ \psi \in \hs \; \mid \; \psi^{\prime \prime
\prime \prime}
\in \hs \ \; {\rm et} \ \; \psi(\pm a) =0 = \psi^{\prime \prime}
(\pm a) \}
\ \ .
\end{equation}
 
Notons que ceci ne repr\'esente qu'une mani\`ere de rendre
auto-adjoint l'op\'erateur
$\ds{\hbar^4 \over 4m^2} \, \ds{d^4 \over dx^4}$ parmi beaucoup
d'autres (d\'etermin\'ees par d'autres conditions aux li\-mites,
par exemple $\psi (\pm a ) = 0 = \psi^{\prime} (\pm a)$),
mais il s'agit de celle qui correspond au syst\`eme physique
que nous consid\'erons.
 
Venons-en maintenant au paradoxe soulev\'e dans notre exemple.
Pour $\psi \in
{\cal D}(H^2) \subset {\cal D}(H)$,
la d\'ecomposition (\ref{h2})
donne
\begin{eqnarray}
\nonumber
\langle H^2 \rangle_{\psi} \ \equiv \ 
\langle \psi , H^2 \psi \rangle 
& \stackrel{(\ref{h2})}{=} & \left\langle \psi ,
\sum_{n=1}^{\infty} E_n^2 \, {\cal P}_n \psi \right\rangle \ = \
\sum_{n=1}^{\infty} E_n^2 \langle \psi , {\cal P}_n \psi \rangle
\\
&= &
\sum_{n=1}^{\infty} E_n^2 \, p_n 
\end{eqnarray}
avec $p_n = |  \langle \varphi_n , \psi \rangle  |^2$.
Si $\psi \in {\cal D} (H)$, nous pouvons
aboutir d'une autre mani\`ere au m\^eme r\'esultat
en utilisant le fait que les projecteurs ${\cal P}_n$ sont auto-adjoints
et orthogonaux
(c'est-\`a-dire ${\cal P}_n {\cal P}_m = \delta_{nm} {\cal P}_n$) :
\begin{eqnarray}
\| H \psi \|^2 & \equiv & \langle H \psi ,  H\psi \rangle
\ = \ 
\left\langle \sum_{n=1}^{\infty} E_n {\cal P}_n \psi ,
\sum_{m=1}^{\infty} E_m {\cal P}_m \psi \right\rangle \ = \
\sum_{n,m=1}^{\infty} E_n E_m \langle {\cal P}_n \psi , {\cal P}_m \psi
\rangle
\nonumber
\\
& = &
\sum_{n,m=1}^{\infty} E_n E_m \langle \psi , {\cal P}_n {\cal P}_m \psi
\rangle
\nonumber
\\
&=&
\sum_{n=1}^{\infty} E_n^2 \, p_n
\ \ .
\end{eqnarray}
La fonction
$\psi(x) = \sqrt{15}/(4a^{5/2}) \,
(a^2 - x^2)$ de notre exemple ne satisfait pas
$\psi^{\prime \prime} (\pm a) =0$ et n'appartient donc pas au domaine de
d\'efinition de $H^2$ : l'expression $\langle \psi, H^2 \psi \rangle$
n'est
donc pas d\'efinie, car la grandeur $H^2$ qui y intervient
n'est pas simplement caract\'eris\'ee par
sa prescription d'op\'eration, mais aussi par son domaine
de d\'efinition. (Autrement dit : quoique l'int\'egrale 
dans l'\'equation (\ref{caf}) soit correctement
\'evalu\'ee,  elle ne peut pas \^etre
identifi\'ee \`a $\langle \psi , H^2 \psi \rangle = \langle H^2
\rangle_{\psi}$ pour la fonction $\psi$ consid\'er\'ee.)
Par contre, nous avons $\psi \in {\cal D} (H)$ et
la valeur moyenne $\langle H^2 \rangle _{\psi}$
peut \^etre \'evalu\'ee
selon
$\sum_{n=1}^{\infty} E_n^2 \, p_n$ ou d'une mani\`ere \'equivalente selon
\[
\| H \psi \|^2 \equiv 
\int_{-a}^{+a} dx \, | (H\psi )(x) | ^2 =
\ds{\hbar^4 \over 4m^2}
\int_{-a}^{+a} dx \, | \psi^{\prime \prime} (x) | ^2 =
\ds{15 \hbar^4 \over 8m^2 a^4}
\ \ .
\]

\bigskip
 
{\bf (4b)}
Consid\'erons maintenant le deuxi\`eme point mentionn\'e dans l'exemple
4,
\`a savoir le spectre de $P$.
Comme indiqu\'e dans l'annexe A.2,
le spectre d'un op\'erateur non auto-adjoint $P$ contient en g\'en\'eral
une partie appel\'ee {\em spectre r\'esiduel} : ce sont tous les
nombres $z \in {\bf C}$ qui ne sont pas valeurs propres de $P$, mais
pour lesquels $\bar z$ est valeur propre de $P^{\dag}$.
Dans l'exemple pr\'esent, le spectre discret et le spectre continu
de $P$ sont vides,
donc le spectre de l'op\'erateur $P$ co\"\i ncide avec son spectre
r\'esiduel.
Comme les fonctions
$\varphi_p (x)=  \, {\rm exp} \, ( \ds{\ri \over \hbar} px )$
avec $p \in \bc$ sont des solutions
de l'\'equation aux valeurs propres pour $P^{\dag}$,
\begin{equation}
\label{spr}
(P^{\dag} \varphi_p)(x) = p \,  \varphi_p(x)
\qquad (p \in {\bf C} \ , \ \varphi_p \in {\cal D}(P^{\dag})
\ , \ \varphi_p \not \equiv 0 )
\ \ ,
\end{equation}
tous les nombres complexes sont valeurs propres de
$P^{\dag}$. En conclusion, le spectre r\'esiduel (et donc le spectre
complet) de $P$ est $\bc$.
$P$ n'\'etant pas auto-adjoint, ce spectre n'a pas d'interpr\'etation
physique directe. Cependant nous allons tout de suite voir 
qu'il contient des informations qui sont importantes pour 
la physique. 
 
Pour \'etudier si le domaine de d\'efinition de $P$ peut \^etre
\'elargi de telle mani\`ere que $P$ devienne auto-adjoint,
il convient d'appliquer la th\'eorie de von Neumann \cite{rs, sg}
selon laquelle il faut \'etudier les valeurs propres complexes
de $P^{\dag}$.
Comme cas particulier de (\ref{spr}), nous avons
\[
P^{\dag} \varphi_{\pm} = \pm \ri \, \varphi_{\pm}
\qquad {\rm avec} \ \ \varphi_{\pm} (x) = {\rm e}^{\mp x / \hbar}
\ \ ,
\]
ou encore
$(P^{\dag} \mp \ri {\bf 1})  \varphi_{\pm} =0$.
Le noyau de l'op\'erateur
$P^{\dag} \pm \ri {\bf 1}$ est donc un espace vectoriel unidimensionnel
:
\begin{eqnarray}
\label{defe1}
n_- (P) & \equiv & {\rm dim \; Ker} \,
(P^{\dag} +  \ri {\bf 1}) =1
\nonumber
\\
n_+ (P) & \equiv & {\rm dim \; Ker} \,
(P^{\dag} -  \ri {\bf 1}) = 1
\ \ .
\end{eqnarray}
Les nombres naturels $n_+ (P)$ et
$n_- (P)$ s'appellent les {\em indices de
d\'efaut de} $P$. 
Leur utilit\'e est montr\'ee par le r\'esultat suivant :  

\newpage 

\begin{theo}[Crit\`ere pour `auto-adjoint']
Soit $A$ un op\'erateur hermitien avec indices de d\'efaut 
$n_+$ et $n_-$. 

\noindent 
{\rm (i)} 
$A$ est auto-adjoint si et seulement si $n_+ = 0= n_-$.
Dans ce cas (et uniquement dans celui-ci), le spectre de $A$ 
est un sous-ensemble de l'axe r\'eel. 

\noindent 
{\rm (ii)}
$A$ admet des extensions auto-adjointes (c'est-\`a-dire 
il est possible de rendre $A$ auto-adjoint en \'elargissant 
son domaine de d\'efinition) si et seulement si $n_+ =  n_-$.
Si $n_+ >0$ et $n_- >0$, le spectre de $A$ est tout le plan complexe.

\noindent 
{\rm (iii)} 
Si l'on a $n_+ =0 \neq n_-$ ou bien $n_- =0 \neq  n_+$,  
l'op\'erateur $A$ n'a pas d'extension  auto-adjointe non-triviale. 
Alors le spectre de $A$ est le demi-plan complexe 
ferm\'e sup\'erieur, respectivement inf\'erieur. 
\end{theo}
Dans le cas $(ii)$, il existe des expressions explicites 
pour les extensions auto-adjointes possibles de $A$ \cite{rs}. 

Dans notre exemple, nous avons $n_+ =  n_- >0$ ; donc 
l'op\'erateur $P$ n'est pas auto-adjoint
et son spectre est tout le plan
complexe (ce que nous savons d\'ej\`a).
Les expressions explicites 
pour les extensions auto-adjointes auxquels nous avons fait allusion,   
impliquent que 
pour tout nombre r\'eel $\alpha$, l'op\'erateur
\begin{equation}
\label{alpha}
P_{\alpha} = \ds{\hbar \over \ri} \, \ds{d \over dx} \quad , \quad
{\cal D} (P_{\alpha}) = \{ \psi \in \hs \, | \,
\psi^{\prime} \in \hs \ \; {\rm et} \ \; \psi(0) =
{\rm e}^{\ri \alpha} \psi(1) \}
\end{equation}
est auto-adjoint 
et on a ${\rm Sp} \, P_{\alpha} = \br$.
 
Du point de vue physique, la condition aux limites
$\psi(0) = {\rm e}^{\ri \alpha} \psi(1)$ veut dire que tout ce qui sort
de l'intervalle $[ 0,1]$ \`a droite rentre de nouveau dans
l'intervalle \`a gauche avec un certain d\'ephasage
(d\'etermin\'e par $\alpha \in \br$) :
ceci permet l'existence d'\'etats
avec une valeur bien d\'efinie de l'impulsion, alors que
la condition aux limites
$\psi(0) = 0 = \psi(1)$ exclut de tels \'etats.
Pour $\alpha =0$, on a des fonctions d'onde p\'eriodiques
et on retrouve l'extension auto-adjointe (\ref{oaa}).

\bigskip
 
{\bf (3b)}
Nous revenons \`a l'affirmation que $\hbar /\ri$ est une valeur
propre de
$A^{\dag}$.
Pour $f \in {\cal D}(A) = \st$, une int\'egration par parties donne
\begin{equation}
\label{surf}
\langle g, A f \rangle \; =  \; \langle \ds{\hbar \over \ri} \left[
(x^3 g)^{\prime}  + 
x^3 g^{\prime} \right] , f \rangle \; + \; 
2 \, \ds{\hbar \over \ri} \left[
x^3 (\bar g  f)(x) \right]_{-\infty}^{+\infty}
\ \ .
\end{equation}
Le terme de surface dans le membre de droite
s'annule si la fonction $g$ ne cro\^it pas plus
vite qu'un polyn\^ome \`a l'infini.
Dans ce cas, l'\'equation pr\'ec\'edente implique
que l'op\'erateur $A^{\dag}$ agit de la m\^eme mani\`ere
que $A$,
\begin{equation}
\label{exa}
A^{\dag} g = \ds{\hbar \over \ri} \, \left[ (x^3 g)^{\prime}
+x^3 g^{\prime} \right]
= \ds{\hbar \over \ri} \, \left[ 3x^2 g
+ 2 x^3 g^{\prime} \right]
\ \ ,
\end{equation}
mais que son domaine de d\'efinition est plus large que $\st$ :
ce domaine contient toutes les fonctions $g$ qui sont telles que
l'expression (\ref{exa}) existe et est de carr\'e sommable.
(Pour toutes ces fonctions, le terme de surface dans l'\'equation
(\ref{surf}) s'annule.)
 
En r\'esum\'e, le domaine de d\'efinition de $A^{\dag}$
est plus grand que celui de $A$ et
l'op\'erateur $A$ n'est donc pas
auto-adjoint. Par ailleurs,
la fonction (\ref{fon})
n'appartient pas \`a ${\cal D}(A)$, mais
elle appartient \`a ${\cal D}
(A^{\dag})$ et $\hbar / \ri$ est donc une valeur propre de $A^{\dag}$.
 
Pour conclure, nous \'etudions bri\`evement
si le domaine de d\'efinition de $A$ peut \^etre
\'elargi de telle mani\`ere que $A$ devienne auto-adjoint. Pour cela
nous faisons de nouveau appel \`a la th\'eorie de von Neumann.
On v\'erifie facilement que
\[
A^{\dag} g_{\pm} = \pm \ri \, g_{\pm}
\qquad {\rm avec} \ \
\left\{
\begin{array}{l}
g_{\pm} (x) =
|x|^{-3/2} \; {\rm exp} \, \left( \pm \ds{1 \over 4 \hbar x^2}
\right)  \quad {\rm pour} \ x \neq 0 \\
g_- (0) = 0 \ \ . 
\end{array}
\right.
\]
Nous avons
$g_- \in {\cal D} (A^{\dag} )$, mais
$g_+ \not\in {\cal D} (A^{\dag} )$ (\`a cause de la croissance
exponentielle de $g_+$ \`a l'origine), donc
\begin{eqnarray*}
n_- (A) & \equiv & {\rm dim \; Ker} \,
(A^{\dag} + \ri {\bf 1}) =1
\\
n_+ (A) & \equiv & {\rm dim \, Ker} \,
(A^{\dag} - \ri {\bf 1}) = 0
\ \ .
\end{eqnarray*}
Du point $(iii)$ du dernier th\'eor\`eme, il s'ensuit maintenant 
qu'il n'y a pas moyen de rendre auto-adjoint l'op\'erateur hermitien
$A$. 

Alors que l'introduction du spectre r\'esiduel appara\^{i}t,  
de premier abord, comme une complication non motiv\'ee
et non physique, les deux derniers exemples
montrent qu'elle est tr\`es int\'eressante du point de la physique.
En effet, pour un op\'erateur $A$, donn\'e sur l'espace de Hilbert, 
il est d'habitude facile de v\'erifier s'il est hermitien 
(en int\'egrant par parties); 
 les indices de d\'efaut de $A$ 
(qui sont \'etroitement li\'es au spectre 
r\'esiduel de $A$) donnent alors une m\'ethode simple et constructive 
pour d\'eterminer toutes les extensions auto-adjointes possibles 
de $A$, c'est-\`a-dire ils d\'ecrivent explicitement toutes les mani\`eres 
pour transformer un op\'erateur hermitien en observable. 
 
Une compr\'ehension plus intuitive des deux derniers exemples
peut \^etre obtenue en consid\'erant les fonctions propres
potentielles des op\'erateurs impliqu\'es et leur
admissibilit\'e pour le probl\`eme physique que l'on \'etudie. 
Pour l'op\'erateur d'impulsion $P$ sur l'intervalle $[0,1]$,
l'onde plane 
 ${\rm exp} \, (\ds{\ri \over \hbar} p x)$ 
 (avec $p \in \br$) r\'esout formellement l'\'equation aux valeurs 
 propres pour $P$, mais elle n'est pas compatible avec les conditions 
 aux limites  $\psi (0) = 0 = \psi (1)$ 
 du puits de potentiel infini; d'un autre c\^ot\'e, 
 la fonction 
 $f_{\lambda} (x) \propto |x|^{-3/2} \, 
 {\rm exp} \, ( \ds{- \ri \lambda \over 4 \hbar x^2})$ 
peut formellement \^etre associ\'ee \`a la valeur propre r\'eelle 
$\lambda$ de $A = PQ^3 +Q^3 P$, 
 mais elle n'est pas de carr\'e sommable \`a cause de 
 son comportement singulier \`a l'origine. 
 Ainsi, les contraintes cruciales pour transformer
 des op\'erateurs hermitiens en observables proviennent, 
 respectivement, des conditions aux limites pour un probl\`eme 
 sur un intervalle compact et de la condition 
 d'int\'egrabilit\'e du carr\'e pour  un probl\`eme 
 sur tout l'espace. 
 (D'ailleurs, ceci sont exactement les m\^emes conditions 
 qui impliquent la quantification des niveaux d'\'energie
sur un intervalle fini et sur tout l'espace, 
respectivement.)


\newpage
 
\begin{thebibliography}{22}
\newcommand{\artref}[5]{{\sc #1} : {\it #2}, #3 {\bf #4} #5}
\newcommand{\bookref}[3]{{\sc #1} : ``{\it #2}$\,$", #3}
\newcommand{\prepref}[3]{{\sc #1} : {\it #2}, #3}
 
\bibitem{d}
\bookref
{P.A.M.Dirac}{The Principles of Quantum Mechanics}{4th edition
(Oxford University Press, Oxford 1958) .}

\bibitem{jau}
\prepref
{J.M.Jauch}{On bras and kets}{in ``Aspects of Quantum Theory",
A.Salam and E.P.Wigner, eds. (Cambridge University Press,
Cambridge 1972) .}

 
\bibitem{grau}
\bookref
{D.Grau}{\"Ubungsaufgaben zur Quantentheorie - Quantentheoretische
Grundlagen}{3.Auflage (C.Hanser Verlag, M\"unchen 1993) .}
 

\bibitem{dieu}
\bookref
{J.Dieudonn\'e}{De la communication entre math\'ematiciens et 
physiciens}{dans ``La Pens\'ee Physique Contemporaine", 
S.Diner, D.Fargue et G.Lochak, \'eds., (\'Editions A.Fresnel, 
Hiersac 1982) .}


\bibitem{ab}
\artref
{M.Breitenecker and H.-R.Gr\"umm}{Remarks on the paper
by P.Bocchieri and A.Loinger ``Nonexistence of the A-B-effect"}{
Nuov.Cim.}{55A}{(1980) 453-455 ;}
 
\artref
{H.-R.Gr\"umm}{Quantum mechanics in a magnetic field}{
Act.Phys.Austr.}{53}{(1981) 113-131 ;}
 
\artref
{S.N.M.Ruisjenaars}{The Aharonov-Bohm
effect and scattering theory}{Ann.Phys.}{
146}{(1983) 1-34 ;}
 
\prepref
{F.Gieres}{\"Uber den Aharonov-Bohm-Effekt}{Diplomarbeit
(Institut f\"ur Theoretische Physik, Universit\"at G\"ottingen, 1983) .}

 
\bibitem{ct}
\bookref
{A.Messiah}{M\'ecanique Quantique, Tome 1 et 2}{(Dunod, Paris 1969) ;}
 
\bookref
{E.Merzbacher}{Quantum Mechanics}{second edition (John
Wiley and Sons, New York 1970) ;}
 
\bookref
{R.P.Feynman, R.B.Leighton and M.Sands}{The Feynman Lectures
on Physics, Vol.3}{(Addison-Wesley, London 1965) ;}
 
\bookref
{K.Gottfried}{Quantum Mechanics}{(Benjamin/Cummings Publ. Co.,
Reading 1966) ;}
 
\bookref
{G.Baym}{Lectures on Quantum Mechanics}{(W.A.Benjamin Inc.,
New York 1969) ;}
 
\bookref
{C.Cohen-Tannoudji, B.Diu et F.Lalo\"e}{
M\'ecanique Quantique, Vol.1 et 2}{deuxi\`eme \'edition
(Hermann, Paris 1977) ;}
 
\bookref
{R.Shankar}{Principles of Quantum Mechanics}{(Plenum, New York 1980) ;}
 
\bookref
{J.-L.Basdevant}{M\'ecanique Quantique}{\'Ecole Polytechnique
(Ellipses, Paris 1986) ;}
 
\bookref
{A.Das and A.C.Melissinos}{Quantum Mechanics - A Modern
Introduction}{(Gordon and Breach Science Publ., New York 1986) ;}
 
\bookref
{Ch.Ng\^o et H.Ng\^o}{Physique Quantique - Introduction avec
Exercices}{(Masson, Paris 1991) ;}
 
\bookref
{P.J.E.Peebles}{Quantum Mechanics}{(Princeton University Press,
Princeton 1992) ;}


\bookref
{F.Schwabl}{Quantum Mechanics}{second revised edition (Springer, Berlin 
1995);}

\bookref
{J.J.Sakurai}{Modern Quantum Mechanics}{revised edition
(Addison-Wesley Publ. Co., Reading 1994) ;}
 
\bookref
{E.Elbaz}{Quantique}{(Ellipses, Paris 1995) .}
 
\bibitem{af}
\bookref
{F.Hirzebruch und W.Scharlau}{Einf\"uhrung in die
Funktional\-analysis}{B.I.-Hochschultaschenb\"ucher
Bd.296 (Bibliographisches Institut,
Mannheim 1971) ;}
 
\bookref
{F.Riesz et B.Sz.Nagy}{Le\c cons d'Analyse
Fonctionnelle}{(Gauthiers-Villars, Paris 1968) ;}
%{F.Riesz and B.Sz.-Nagy}{Functional Analysis}{(Frederick Ungar
%Publ. Co., New York 1955) ;}
 
\bookref
{N.I.Akhiezer and I.M.Glazman}{Theory of Linear Operators in
Hilbert Space, Vol.I and II}{(Frederick Ungar Publ. Co., New York 1961
and 1963) ;}
 
\bookref
{N.Dunford and J.T.Schwartz}{Linear Operators,
Vol.I-III}{(Interscience Publishers, New York 1958, 1963 and 1971) ;}
 
\bookref
{J.Weidmann}{Lineare Operatoren in Hilbertr\"aumen}{(B.G.Teubner,
Stuttgart 1976) .}

 
 
\bibitem{rs}
\bookref
{M.Reed and B.Simon}{Methods of Modern Mathematical Physics, Vol.1-
Functional Analysis}{revised edition
(Academic Press, New York 1980) ;}
 
\bookref
{M.Reed and B.Simon}{Methods of Modern Mathematical Physics, Vol.2-
Fourier Analysis, Self-Adjointness}{(Academic Press, New York 1975) .}
 
   
\bibitem{sg}
\bookref
{S.Gro\ss mann}{Funktionalanalysis I, II - im Hinblick auf
Anwendungen in der Physik}{Studientext (Akademische
Verlagsgesellschaft, Wiesbaden 1975, 1977).}

 
\bibitem{bgc}
\bookref
{P.Benoist-Gueutal et M.Courbage}{Math\'ematiques pour
la Physique, Tome 3 - Op\'erateurs lin\'eaires dans les espaces
de Hilbert}{(Eyrolles, Paris 1993) .}
 

\bibitem{ri}
\bookref
{R.D.Richtmyer}{Principles of Advanced Mathematical Physics I}{
(Springer Verlag, Berlin 1978) .}
 
 
\bibitem{lio}
\bookref
{E.Weislinger}{Math\'ematiques pour Physiciens
(2\`eme et 3\`eme cycles, avec rappels de 1er
cycle)}{(Ellipses, Paris 1991) ;}
 
\bookref
{R.Dautray et J.-L.Lions}{Analyse Math\'ematique et Calcul
Num\'erique pour les Sciences et les Techniques,
Tomes 1-3}{Collection CEA (Masson, Paris 1985) ;}
 
\bookref
{R.Geroch}{Mathematical Physics}{(The University of Chicago Press,
Chicago 1985) ;}
 
\bookref
{H.Triebel}{H\"ohere Analysis}{2.Auflage (Verlag Harri Deutsch, Thun
1980) ;}


\bookref
{Ph.Blanchard and E.Br\"uning}{Distributionen und 
Hilbertraumoperatoren - Mathematische Methoden der 
Physik}{(Springer Verlag, Berlin 1993) ;} 

\bookref
{E.Zeidler}{Applied Functional Analysis - Applications to Mathematical
Physics}{Appl.Math.Sci. Vol.108 (Springer Verlag, Berlin 1995) ;}

\bookref
{M.A.Shubin (ed.)}{Partial Differential Equations VII - 
Spectral Theory of Differential Operators}{Encyclopaedia
of Mathematical Sciences, Vol. 64, (Springer Verlag, Berlin 1994) .} 
 
\bibitem{tj}
\bookref
{T.F.Jordan}{Linear Operators for Quantum Mechanics}{(John
Wiley and Sons, New York 1969) .}
 
\bibitem{krey}
\bookref
{E. Kreyszig}{Introductory Functional Analysis with Applications}{Wiley
Classics Library Edition (John Wiley, New York 1989) .}
 
 
\bibitem{fh}
\bookref
{F.Hund}{Geschichte der Physikalischen Begriffe}{
B.I.-Hochschultaschen\-b\"ucher Bd.543, (Bibliographisches
Institut, Mannheim 1968) .}
 
\bibitem{jvn}
\prepref
{J.von Neumann}{``Mathematische Grundlagen der Quantenmechanik"}{
(Springer Verlag, Berlin 1932); English transl. : {\it
``Mathematical Foundations of Quantum Mechanics"}
(Princeton University Press, Princeton 1955) .}
 
\bibitem{ch}
\bookref
{R.Courant and D.Hilbert}{Methoden der Mathematischen Physik,
Bd.1 und 2}{(Springer Verlag, Berlin 1924 and 1937), (English
transl. : {\it ``Methods of Mathematical Physics, Vol.1 and 2"}
(Interscience Publ., New York 1966 and 1962) .}
 
\bibitem{cr}
\bookref
{C.Reid}{Hilbert - Courant}{(Springer Verlag, New York 1986) .}
 
\bibitem{ls}
\bookref
{L.Schwartz}{Th\'eorie des Distributions}{(Hermann, Paris 1966) .}
 
\bibitem{gv}
\bookref
{I.M.Gel'fand and N.Ya.Vilenkin}{Les Distributions, Vol.4 -
Applications de l'Analyse Harmonique}{(Dunod, Paris 1967) .}
%{I.M.Gel'fand and N.Ya.Vilenkin}{Generalized Functions Vol.4 -
%Applications of Harmonic Analysis}{
%(Academic Press, New York 1964) .}
 
\bibitem{vdw}
\bookref
{B.L.van der Waerden}{Sources of Quantum Mechanics}{edited with
a historical introduction (North-Holland Publ. Co., Amsterdam 1967) .}
 
\bibitem{rob}
\bookref
{D.W.Robinson}{The Thermodynamic Pressure in Quantum Statistical
Mecha\-nics}{Lecture Notes in Physics 9,
(Springer Verlag, Berlin 1971).}
 
\bibitem{bs}
\bookref
{H.L.Cycon, R.G.Froese, W.Kirsch and B.Simon}{Schr\"odinger
Ope\-rators - with Applications to Quantum Mechanics and Global
Geometry}{(Springer Verlag, Berlin 1987) ;}
 
\bookref
{R.Carmona and J.Lacroix}{Spectral Theory of Random Schr\"odinger
Operators}{(Birkh\"auser, Boston 1990) ;}
 
\bookref
{S.Albeverio, F.Gesztesy, R.Hoegh-Krohn and H.Holden}{Sol\-vable
Models in Quantum Mechanics}{Texts and Monographs in Physics,
(Springer Verlag, Berlin 1988) .}
 
\bibitem{ber}
\bookref
{F.A.Berezin and M.A.Shubin}{The Schr\"odinger Equation}{Mathematics
and its Applications Vol.66 (Kluwer Academic Publ., Dordrecht 1991).}
 
\bibitem{th}
\bookref
{W.Thirring}{A
Course in Mathematical Physics, Vol.3 -
Quantum Mechanics of Atoms and Molecules}{second edition
(Springer Verlag, Berlin 1991) .}
 
\bibitem{aw}
\prepref
{A.S.Wightman}{Introduction to some aspects of the relativistic
dynamics}{in ``High Energy Electromagnetic Interactions and
Field Theory", M.L\'evy ed. (Gordon an Breach, New York 1967) .}
 
\bibitem{amj}
\bookref
{W.O.Amrein, J.M.Jauch and K.B.Sinha}{Scattering Theory in
Quantum Mechanics}{Lecture Notes and Supplements in Physics Vol.16
(Benjamin, Reading 1977) .}
 
\bibitem{au}
\bookref
{J.Audretsch und K.Mainzer (Hrsg.)}{Wieviele Leben hat Schr\"odingers
Katze?}{(Bibliographisches Institut, Mannheim 1990) ;}
 
\prepref
{A.Shimony}{Les fondements conceptuels de la m\'ecanique
quantique}{dans ``La Nouvelle Physique", P.Davies ed. (Flammarion,
Paris 1993) ;}
%{A.Shimony}{Conceptual foundations of quantum
%mechanics}{in ``The New Physics", P.Davies ed. (Cambdridge
%University Press, Cambridge 1993) ;}
 
\bookref
{J.A.Wheeler and W.H.Zurek (Eds.)}{Quantum Theory and
Measurement}{(Princeton University Press, Princeton 1983) .}
 
\bibitem{pri}
\bookref
{I.Prigogine}{Les Lois du Chaos}{Nouvelle Biblioth\`eque
Scientifique (Flammarion, Paris 1994) .}
 
\bibitem{fano}
\bookref
{G.Fano}{Mathematical Methods of Quantum Mechanics}{(McGraw-Hill
Book Co., New York 1971) .}
 
\bibitem{jr}
\artref
{J.E.Roberts}{The Dirac Bra and Ket Formalism}{J.Math.Phys.}{7}{
(1966) 1097-1104 ;}
 
\artref
{J.E.Roberts}{Rigged Hilbert spaces in quantum
mechanics}{Commun.Math.Phys.}{3}{(1966) 98-119 .}
 
\bibitem{jpa}
\artref
{J.-P.Antoine}{Dirac formalism and symmetry problems in quantum
mechanics I : General
Dirac formalism}{J.Math.Phys.}{10}{(1969) 53-69 ;}
 
\artref
{O.Melsheimer}{Rigged Hilbert space formalism as an extended
mathematical formalism for quantum systems. I. General theory}{
J.Math.Phys.}{15}{(1974) 902-916 ;}
 
\artref
{O.Melsheimer}{Rigged Hilbert space formalism as an extended
mathematical formalism for quantum systems. II.
Transformation theory in nonrelativistic quantum mechanics}{
J.Math.Phys.}{15}{(1974) 917-925 ;}

\bookref
{S.J.L.van Eijndhoven and J.de Graaf}{A Mathematical Introduction 
to Dirac's Formalism}{(North-Holland, Amsterdam 1986) .} 


\bibitem{gl}
\bookref
{G.Ludwig}{Foundations of Quantum Mechanics, Vol.1 and 2}{Texts and
Monographs in Physics (Springer
Verlag, Berlin 1983 and 1985) .}
 
   
\bibitem{gap}
\bookref
{A.Galindo and P.Pascual}{Quantum Mechanics, Vol.1 and 2}{(Springer
Verlag, Berlin 1990 and 1991) ;}
 
\bookref
{G.Grawert}{Quantenmechanik}{3.Auflage
(Akademische Verlagsgesellschaft,
Wiesbaden 1977) ;}
 
\bookref
{G.C.Hegerfeldt}{Quantenmechanik}{Skriptum
1974/75 und 1981/82 (Institut f\"ur Theoretische Physik,
Universit\"at G\"ottingen) .}

 
\bibitem{ll}
\bookref
{L.D.Landau and E.M.Lifschitz}{M\'ecanique Quantique}{(Editions
MIR, Moscou 1967) ;}
 
\bookref
{L.I.Schiff}{Quantum Mechanics}{3rd edition
(McGraw-Hill, New York 1968) ;}
 
\bookref
{D.S.Saxon}{Elementary Quantum Mechanics}{(Holden-Day, San Francisco
1968) ;}
 
\bookref
{S.Gasiorowicz}{Quantum Physics}{(John Wiley and Sons, New York
1974) .}
 
\bibitem{pd}
\bookref
{H.Haken and H.C.Wolf}{The Physics of Atoms and Quanta - Introduction
to Experiments and Theory}{4th edition
(Springer Verlag, Berlin 1994) ;}
 
\bookref
{P.C.W.Davies}{Quantum Mechanics}{(Chapman and Hall, London 1984) ;}
 
\bookref
{B.H.Bransden and C.J.Joachain}{Introduction to Quantum
Mechanics}{(Longman Scientific and Technical, Essex 1989) ;}
 
\bookref
{F.Mandl}{Quantum Mechanics}{(John Wiley and Sons, New York 1992);}

\bookref
{I.Bialynicki-Birula, M.Cieplak and J.Kaminski}{Theory of Quanta}{(Oxford 
University Press, Oxford 1992) .}


\bibitem{bal}
\bookref
{L.E.Ballentine}{Quantum Mechanics - A Modern Development}{(World
Scientific, Singapore 1998);}
 
\bookref
{A.Bohm}{Quantum Mechanics : Foundations and Applications}{Second
edition (Springer Verlag, Berlin 1986) .}
 
\bibitem{jmj}
\bookref
{C.Piron}{M\'ecanique Quantique - Bases et Applications}{(Presses
polytechniques et universitaires romandes, Lausanne 1990) ;}
 
\bookref
{J.M.Jauch}{Foundations of Quantum Mechanics}{(Addison-Wesley Publ.,
Reading 1968) .}
 
\bibitem{rj}
\bookref
{R.Jost}{Quantenmechanik I,II}{(Verlag des Vereins der Mathematiker
und Physiker an der ETH Z\"urich, Z\"urich 1969) ;}
 
%\bibitem{am}
\bookref
{W.O.Amrein}{Non-Relativistic Quantum Dynamics}{Mathematical
Physics Studies Vol.2 (D.Reidel Publ. Co., Dordrecht 1981) .}
 
\bibitem{blt}
\bookref
{N.N.Bogolubov, A.A.Logunov and I.T.Todorov}{Introduction to
Axiomatic Quantum Field Theory}{Mathematical Physics Monograph
Series Vol.18 (Benjamin/Cummings Publ. Co., Reading 1975) .}
 

\bibitem{go}
\bookref
{B.R.Gelbaum and J.M.H.Olmsted}{Counterexamples in
Analysis}{(Holden-Day, San Francisco 1964) .}


\bibitem{car}
\artref
{P.Carruthers and M.M.Nieto}{Phase and angle variables in quantum
mechanics}{Rev.Mod.Phys.}{40}{(1968) 411-440 .}


\bibitem{ju}
\artref
{D.Judge}{On the uncertainty relation for $L_z$ and $\varphi$}{
Phys.Lett.}{5}{(1963) 189 .}


\bibitem{krau}
\artref
{K.Kraus}{Remark on the uncertainty between angle and angular momentum}{
Z.Phys.}{188}{(1965) 374-377 .}
 

 
 
\end{thebibliography}
\end{document}